\documentclass[aps, prx, reprint, twocolumn, superscriptaddress, floatfix]{revtex4-2}
\usepackage{graphicx} % Required for inserting images
\usepackage{upgreek}
\usepackage{amsmath}
\usepackage{braket}
\usepackage{float}
\usepackage{hyperref}
\usepackage{natbib}  % Use this for citation styles
\usepackage{balance}
\floatstyle{plain}
\usepackage{xcolor}

\usepackage{times}
\usepackage{ulem}
\usepackage{ulem}
\usepackage{epsfig}
\usepackage{epstopdf}

\usepackage{booktabs}
\usepackage{indentfirst}
\usepackage{amsmath}

\providecommand{\ignore}[1]{}

\newcommand{\expect}[1]{\left<{#1}\right>}

\begin{document}

\title{Statistics of Strongly Coupled Defects in Superconducting Qubits}
\author{S. Weeden}
\email{scweeden@wisc.edu}
\affiliation{Department of Physics, University of Wisconsin-Madison, Madison, Wisconsin 53706, USA}
\author{D. C. Harrison}
\affiliation{Department of Physics, University of Wisconsin-Madison, Madison, Wisconsin 53706, USA}
\affiliation{Qolab, Madison, Wisconsin 53706, USA}
\author{S. Patel}
\affiliation{Department of Physics, University of Wisconsin-Madison, Madison, Wisconsin 53706, USA}
\author{M. Snyder}
\affiliation{Department of Physics, University of Wisconsin-Madison, Madison, Wisconsin 53706, USA}
\author{E. J. Blackwell}
\affiliation{Department of Physics, University of Wisconsin-Madison, Madison, Wisconsin 53706, USA}
\author{G. Spahn}
\affiliation{Department of Physics, University of Wisconsin-Madison, Madison, Wisconsin 53706, USA}
\author{S. Abdullah}
\affiliation{Department of Physics, University of Wisconsin-Madison, Madison, Wisconsin 53706, USA}
\author{Y. Takeda}
\affiliation{Department of Physics, University of Wisconsin-Madison, Madison, Wisconsin 53706, USA}
\author{B. L. T. Plourde}
\affiliation{Department of Physics, University of Wisconsin-Madison, Madison, Wisconsin 53706, USA}
\affiliation{Qolab, Madison, Wisconsin 53706, USA}
\author{J. M. Martinis}
\affiliation{Qolab, Madison, Wisconsin 53706, USA}
\author{R. McDermott}
\email{rfmcdermott@wisc.edu}
\affiliation{Department of Physics, University of Wisconsin-Madison, Madison, Wisconsin 53706, USA}
\affiliation{Qolab, Madison, Wisconsin 53706, USA}

\begin{abstract}

Decoherence in superconducting qubits is dominated by defects that reside at amorphous interfaces. Interaction with discrete defects results in dropouts that complicate qubit operation and lead to nongaussian tails in the distribution of qubit energy relaxation time $T_1$ that degrade system performance. Spectral diffusion of defects over time leads to fluctuations in $T_1$, posing a challenge for calibration. In this work, we measure the energy relaxation of flux-tunable transmons over a range of operating frequencies. We vary qubit geometry to change the interface participation ratio by more than an order of magnitude. Our results are consistent with loss dominated by discrete interfacial defects. Moreover, we are able to localize the dominant defects to within 500~nm of the qubit junctions, where residues from liftoff are present. These results motivate new approaches to qubit junction fabrication that avoid the residues intrinsic to the liftoff process.

\end{abstract}

\maketitle
\section{Introduction}

State-of-the-art superconducting qubits achieve entangling gate infidelity approaching $10^{-3}$ \cite{Riken2024, GoogleQuantum2024, Gao2025}, limited by energy relaxation from coupling to two-level state (TLS) defects that reside at amorphous interfaces \cite{Schickfus1977, Martinis2005, Macha2010, Lisenfeld2019, Molina-Ruiz2021}. In the continuum limit, coupling of resonant defects to electric fields in the qubit can be modeled as a nonzero loss tangent $\tan \delta$. At the same time, interaction of the qubit with single TLS results in discrete ``dropouts" in the qubit spectrum characterized by strong suppression of qubit energy relaxation time $T_1$. For fixed-frequency qubits, strong coupling to individual defects renders some fraction of the qubits unusable, reducing the yield of large-scale quantum processors \cite{Osman2023}. For frequency-tunable qubits, the need to navigate around TLS dropouts makes processor calibration extremely challenging, especially as the TLS fluctuate in time \cite{Lisenfeld2015, Muller2015, Klimov2018, Bylander2019, Weides2019}. Continued progress in the
development of multiqubit arrays that aspire to fault tolerance will require a deep understanding of qubit-TLS physics and suppression of the density of strongly coupled defects. 

In this paper, we examine the frequency- and time-dependent coupling of superconducting transmon qubits to individual dielectric TLS defects. We use the technique of swap spectroscopy to map out the spectrum of strongly coupled defects over a wide range of qubit operating frequency \cite{Barends2013}. In addition, we monitor fluctuations of the defect bath over time. We study qubit-TLS coupling over a range of qubit geometry, and we find scaling consistent with a high density of TLS defects that reside at device interfaces. Strong coupling to discrete TLS leads to a broad distribution of qubit energy relaxation times; the low-$T_1$ tails of the distribution are known to limit performance of large-scale processors \cite{positionPaper2024}. We find that optimized processing and qubit geometry yield tight distributions, compatible with high-performance multiqubit processors. In particular, it is critical to minimize the participation ratio of structures fabricated with additive liftoff processing, which leaves lossy residues that suppress qubit $T_1$.
\begin{figure}[H]
    \centering
    \includegraphics[width=\columnwidth]{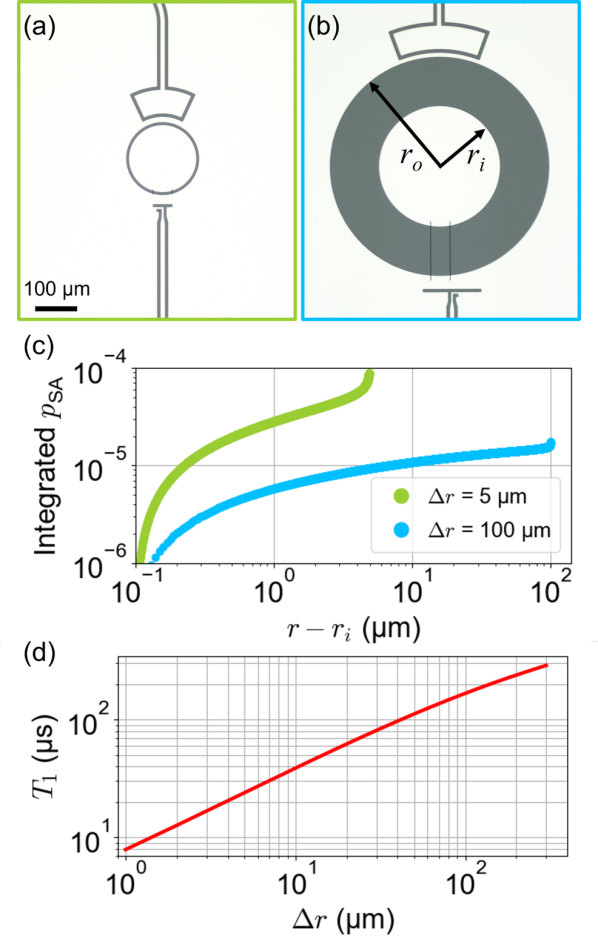}
    \caption{Optical micrograph of single-ended long-liftoff transmon qubits with circular island and gap $\Delta r$ to ground equal to (a) 5~$\upmu$m and (b) 100~$\upmu$m. (c) Integrated substrate-air (SA) interface participation $p_\text{SA}$ \textit{versus} separation $r-r_i$ from the qubit island. (d) Calculated $T_1$ due to TLS as a function of $\Delta r$ for fixed $E_\mathrm{c} / h$~=~270~MHz, assuming  defect density $\sigma$ = 2/GHz/$\upmu$m$^2$ at the SA interface and maximum TLS dipole moment $p_{\text{max}}$ = 5~Debye. For fixed $\sigma p_\text{max}^2$, qubit energy relaxation time scales approximately as $T_1 \propto \Delta r^{0.6}$.}
    \label{fig:Figure1}
\end{figure}

\section{Experiment}

We probe energy relaxation in twelve aluminum-on-silicon devices with six different geometries; see Fig.~\ref{fig:Figure1}(a-b). All qubits are single-ended charge-insensitive transmons with flux-tunable compound Al/AlO$_x$/Al junctions to allow \textit{in situ} tuning of the Josephson energy $E_{\rm J}$. The qubit island is realized as a circular patch with radius $r_i$ embedded in a cavity in the groundplane of the circuit with radius $r_o$. In order to probe sensitivity of qubit energy relaxation to interfacial defects, we vary the gap $\Delta r \equiv r_o - r_i$ from qubit island to ground while maintaining a fixed qubit charging energy $E_\mathrm{C}/h \sim$~270~MHz. For all devices, the maximum qubit operating frequency is in the range 5.4-5.6~GHz, corresponding to a ratio of Josephson energy to charging energy $E_\text{J}/E_\text{C}$ around 60. We expect loss from defects at the substrate-air (SA) interface in the gap between qubit island and ground to scale linearly with the SA participation ratio, defined as the ratio of electric field energy stored in the SA interface to total electric field energy in the device. In Fig. \ref{fig:Figure1}(c) we plot the integrated SA participation ratio for gap sizes of 5 and 100 $\upmu$m. As electric field scales roughly as $\Delta r^{-1}$ and the number of defects is linear in $\Delta r$ for fixed defect density, we expect qubit relaxation time to increase roughly linearly with gap $\Delta r$ from qubit island to ground. A detailed calculation predicts the scaling $T_1 \propto \Delta r^{0.6}$; see Fig.~\ref{fig:Figure1}(d) and Appendix~D. Here, we study qubits with gap sizes of 5, 20, and 100~$\upmu$m. For each gap variant, we examine two different junction lead geometries. Half of the devices are ``long-liftoff" qubits, in which the entirety of the junction lead is formed in the double-angle evaporation and liftoff step used to create the Josephson junction. The other half are ``short-liftoff" qubits, where most of the lead is defined during the initial optical lithography and etch step that creates the qubit island and groundplane. For these devices, the liftoff portion of the junction leads is only 2~$\upmu$m long, independent of gap size. See Fig. \ref{fig:sup1} in Appendix \ref{appendixA} for details. All measurements are performed in a dilution refrigerator at mixing chamber temperatures of 10-15 mK.

\begin{figure}
    \includegraphics[width=\columnwidth]{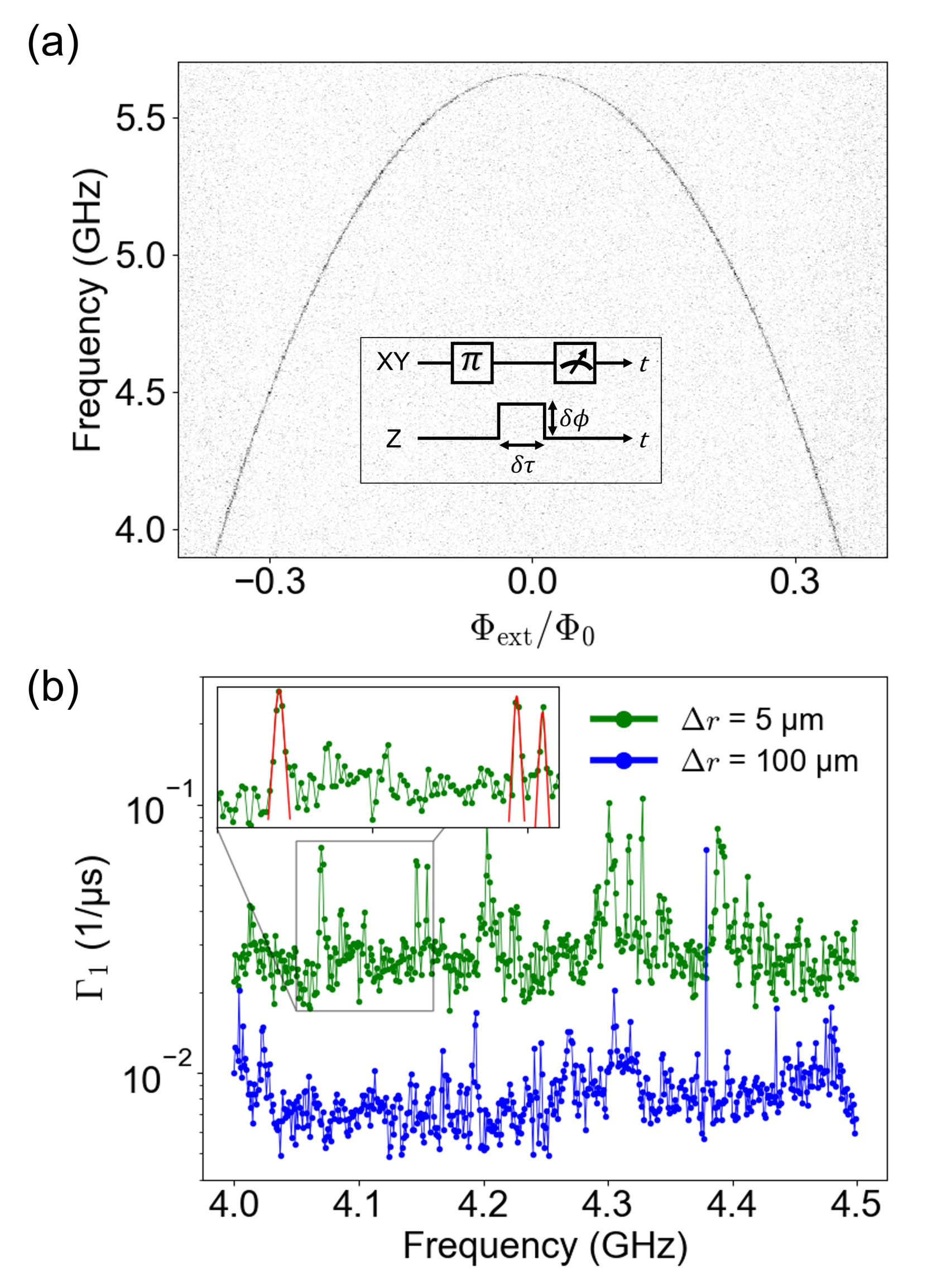}
    \caption{(a) Flux dispersion of a 100 $\upmu$m-gap short-liftoff qubit. The timing diagram of the swap sequence is shown in the inset. (b) Measured $\Gamma_1$ \textit{vs}. operating frequency for short-liftoff qubits with gaps of 5 and 100 $\upmu$m. The inset shows example fits obtained using Eq.~\ref{eq:lorentzian}.}
    \label{fig:f2}
\end{figure}

To probe variations in qubit relaxation rate versus qubit operating frequency, we use swap spectroscopy \cite{Barends2013}; see Fig.~\ref{fig:f2}(a). Here, we bias the qubit to the flux-insensitive upper sweet spot and apply a $\pi$-pulse to prepare the qubit $\ket{1}$ state. We then apply a flux pulse to rapidly tune the qubit to a different operating point where it idles for a variable time before returning to the upper sweet spot for measurement. To account for slow drifts in qubit frequency due to changes in the local magnetic flux environment, we perform qubit spectroscopy at multiple bias points before each swap scan and feed back to stabilize the qubit operating point; these measurements are described in detail in Appendix \ref{appendixB}. For an idle point corresponding to qubit operating frequency $f$ that is near the resonant frequency $f_{\rm d}$ of a TLS defect,  interaction of the qubit with the TLS leads to population decay and enhanced qubit energy relaxation rate $\Gamma_1$. In the relevant incoherent limit of qubit-TLS coupling much less than the defect relaxation rate $g \ll \Gamma_{\rm d}$, the relaxation rate of the qubit is obtained from Fermi's Golden Rule, where we use a normalized Lorentzian lineshape to model the TLS defect \cite{tlsDephasing}: 
\begin{equation}
    \Gamma_1 = \frac{2g^2 \Gamma_\text{d}}{\Gamma_\text{d}^2 + 4\pi^2(f - f_\text{d})^2}.
    \label{eq:lorentzian}
\end{equation}

In Fig. \ref{fig:f2}(b), we show representative linecuts of $\Gamma_1$ as a function of qubit operating frequency. Qubit relaxation rate varies strongly with operating point, with excursions around the mean of order 20\%. Strongly-coupled defects appear as peaks in the data. These peaks are fit with Eq. \ref{eq:lorentzian} to yield both qubit-TLS coupling $g$ and TLS relaxation rate $\Gamma_{\rm d}$.

\begin{figure*}
    \centering
    \includegraphics[width=0.95\linewidth]{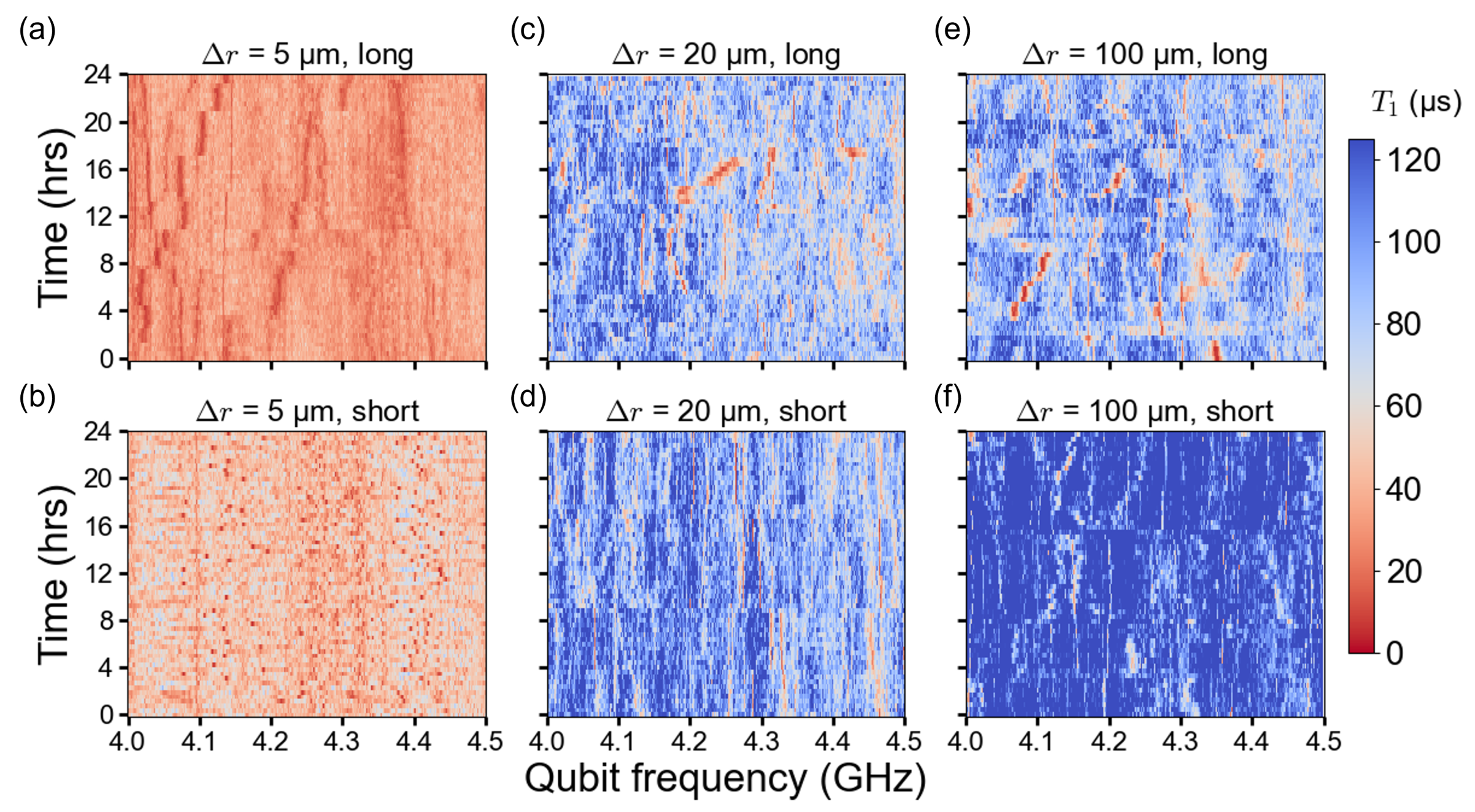}
    \caption{Repeated swap spectroscopy scans for the six device geometries. Qubit gap $\Delta r$ increases from left to right; the top row corresponds to long-liftoff qubits, while the bottom row corresponds to short-liftoff qubits. The extracted mean relaxation times with standard deviation are (a) 32 $\pm$ 7 $\upmu$s, (b) 45 $\pm$ 12 $\upmu$s, (c) 84 $\pm$ 23 $\upmu$s, (d) 95 $\pm$ 26 $\upmu$s, (e) 85 $\pm$ 24 $\upmu$s, and (f) 132 $\pm$ 34  $\upmu$s.}

    \label{fig:f3} 
    
\end{figure*}

The pattern of qubit relaxation rate \textit{versus} frequency constitutes a defect ``fingerprint" that reflects the instantaneous distribution in frequency of TLS defects that couple strongly to zero-point fields in the qubit. However, these defects also fluctuate in time. We study temporal fluctuations in qubit $T_1$ by repeating the swap experiment many times over the course of a full day. In Fig. \ref{fig:f3}, we show swap data for the six qubit geometries considered here. Each plot comprises 24 hours of repeated swap scans, a duration that captures TLS frequency fluctuations occurring on timescales of minutes to hours. In all plots, the frequency resolution is 1~MHz and the timing resolution is 20 minutes. The gap from qubit island to ground increases from left to right; the upper panels correspond to long-liftoff qubits, while the lower panels correspond to short-liftoff qubits. The temporal variation of $T_1$ at a fixed operating frequency results from spectral diffusion of TLS defects due to  interaction of high-frequency TLS with a bath of low-frequency
thermal fluctuators (TF) \cite{Muller2015, Klimov2018}. In this picture, TF fluctuate on timescales set by phonon excitation and relaxation rates \cite{Phillips1981, Faoro20151}.  Strong coupling of TLS to a single dominant TF gives rise to telegraph noise in the TLS frequency. In the continuum limit of coupling to many TF, spectral diffusion emerges. Qualitative analysis of the data in Fig. \ref{fig:f3} suggests that both phenomena are present. We present a more detailed analysis of TLS spectral diffusion in Appendix \ref{appendixE}.

In Fig.~\ref{fig:f4_T1Hist}, we plot cumulative distributions of $T_1$ for each device geometry; each trace represents data taken across the band from 4.0-4.5~GHz over 24 hours for two qubits with identical geometry. For qubits with larger gap $\Delta r$ from island to ground we see longer $T_1$ times.  
For a fixed gap, the median $T_1$ times are generally higher for short-liftoff qubits than for long-liftoff qubits; the long-liftoff 100 $\upmu$m-gap qubits show especially pronounced low-$T_1$ tails. Both of these effects are well-modeled by assuming an enhanced density of defects at the edges of the junction leads; see Appendix \ref{appendixC} for details.
The mitigation of such dropouts is critical for the design of robust error-corrected quantum processors \cite{positionPaper2024}; while the surface code is tolerant to dropouts at the 1\% level \cite{Fowler2012, FowlerMartinis2012}, it is expected that fidelity of the system as a whole will be limited by the worst-performing qubits for a higher density of $T_1$ dropouts. 

\begin{figure}[b]
    \centering    \includegraphics[width=\columnwidth]{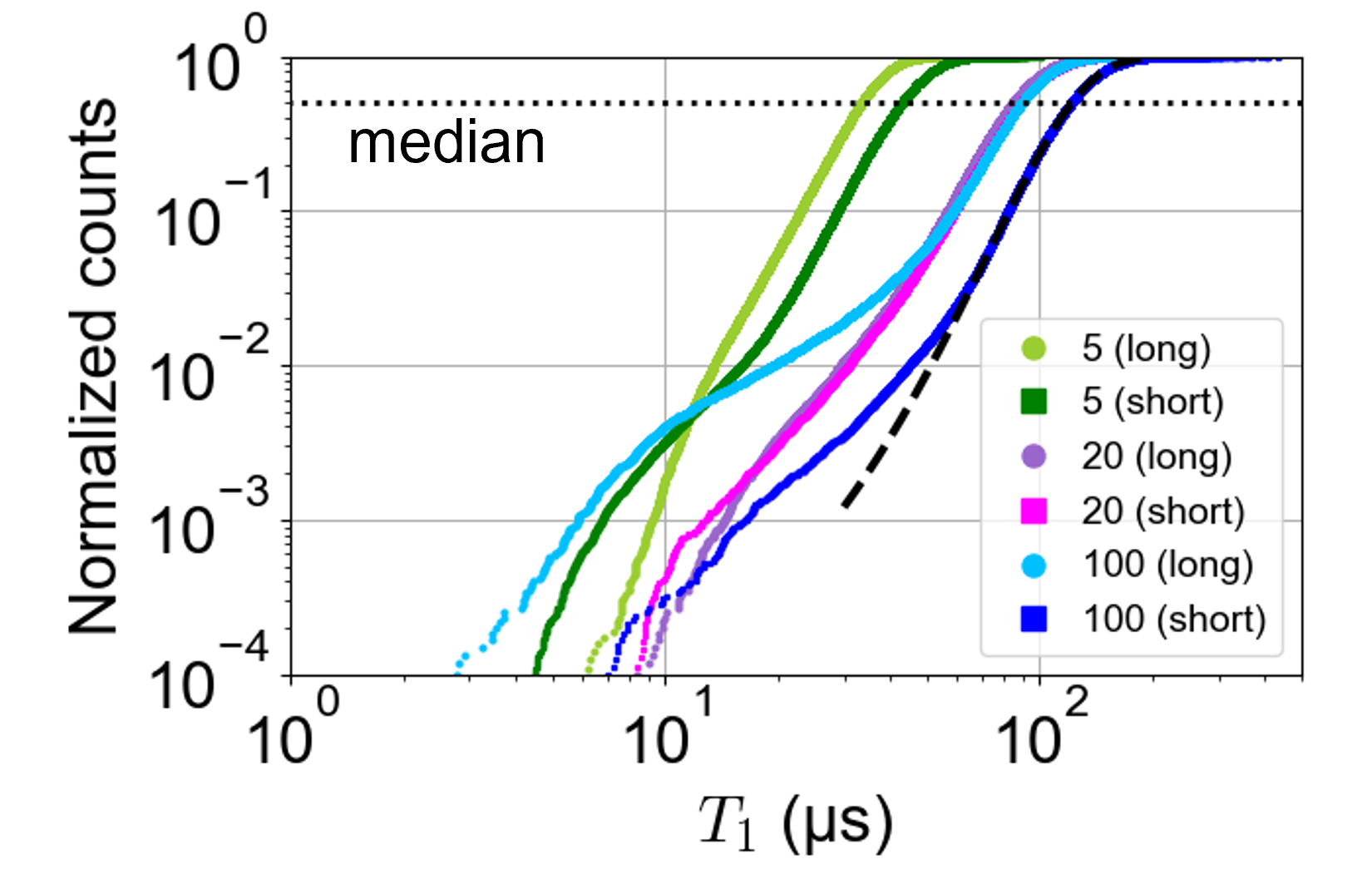}
    \caption{Integrated $T_1$ histograms for all 12 devices measured, with data from nominally identical qubits combined into a single trace. Each curve contains around 80,000 $T_1$ measurements acquired over the 24-hour experiment. The black dotted line passes through the median $T_1$ of each dataset. The distribution of the short-liftoff 100~$\upmu$m-gap qubits is well-modeled by a gaussian (dashed black line).}
    \label{fig:f4_T1Hist} 
\end{figure}

From our fits to the dominant $T_1$ dropouts, we extract distributions of TLS-qubit coupling strengths $g$ and TLS relaxation rate $\Gamma_{\rm d}$; details of the analysis are presented in Appendix \ref{appendixB}.
We find coupling strengths $g/2\pi$ extending to 0.3~MHz, with defect relaxation rate $\Gamma_{\rm d}$  in the range 2-30~$\upmu$s$^{-1}$. These coupling strengths and rates are consistent with previous reports in the literature \cite{Katz2010, Barends2013}. The upper range of measured coupling $g$ is compatible with defects at the edges of the junction leads within 500~nm of the junction itself; see Appendix C below for full analysis. For defects inside the tunnel barrier itself, where the zero-point electric fields are of order kV/m, we expect coupling strengths $g/2\pi$ of order tens of MHz. Such dominant defects are statistically avoided due to the small junction area (nominally 0.04~$\upmu$m$^2$).

\section{Modeling}

To model the data, we consider a bath of TLS defects that couple to the zero-point electric fields $\mathbf{E}$ in the qubit via an electric dipole moment $\mathbf{p}$. The energy of this interaction is expressed as $\hbar g=\mathbf{E}\cdot\mathbf{p}$.  We use the standard tunneling model to describe the TLS defects \cite{Phillips1981}. Here, states $\ket{L}$ and $\ket{R}$ are associated with distinct spatial charge configurations that differ in energy by $\Delta$ and are coupled by a tunneling matrix element $\Delta_0$; the defect dipole moment arises from charge tunneling between these configurations. The defect energy eigenstates can be written as
$\ket{g} = \cos\left({\theta}/{2}\right) \ket{L} + \sin\left({\theta}/{2}\right) \ket{R}$
and $\ket{e} =\sin\left({\theta}/{2}\right) \ket{L} - \cos\left({\theta}/{2}\right) \ket{R}$,
where we have introduced the TLS angle $\theta \equiv \arctan(\Delta/\Delta_0)$. The defect energies are $\epsilon = \pm \sqrt{\Delta^2 + \Delta_0^2}$. The defect dipole moment has magnitude $p = p_{\rm max} \sin\theta$, and the coupling of the TLS to the qubit is given by $\hbar g = E p \cos \eta$, where $E$ is the magnitude of the local electric field and $\eta$ is the real-space angle between $\mathbf{E}$ and $\mathbf{p}$.

\begin{figure}[b]
    \centering
    \includegraphics[width=\columnwidth, keepaspectratio]{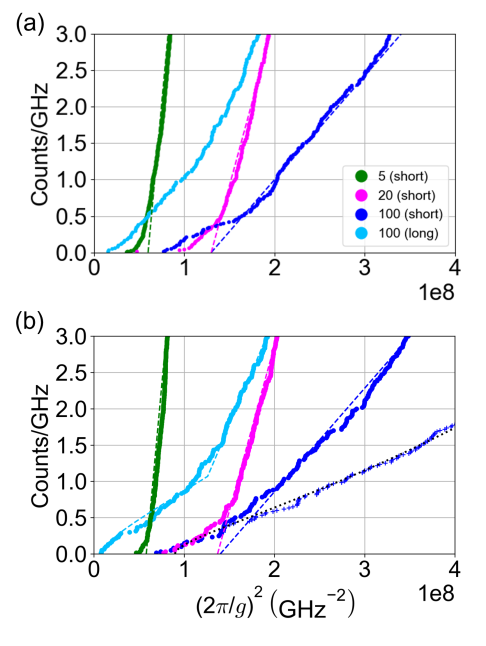}
    
    \caption{Cumulative histograms of $(2\pi/g)^2$ for (a) measured data and (b) Monte Carlo simulations. The green, pink, and dark blue dashed lines are generated using Eq. \ref{eq:slope} with $p_\text{max} = $ 5~Debye and $\sigma = $ 2/GHz/$\upmu$m$^2$. In (b), the simulated lead defects for the 100 $\upmu$m short-liftoff qubits are plotted separately as the dark blue `+' symbols. The dotted line is obtained from an analytical expression for lead defects with linear density $\lambda = 0.4$/GHz/$\upmu$m. In (b), alongside Monte Carlo simulations of the long-lead device (light blue points), we plot the sum of the analytic expressions for lead and gap defects (dashed light blue line).  From our analysis, we find that the most strongly-coupled defects at low $\left(2 \pi/g\right)^2$ are located within 500~nm of the Josephson junction. See Appendix C for details.}
    \label{fig:f5}
\end{figure}

The standard tunneling model assumes a defect density of states that is independent of $\Delta$ and proportional to $1/\Delta_0$. Performing the average over TLS real-space angles $\eta$, we find a distribution of dipole moments given by \cite{Martinis2005}

\begin{equation}
    \frac{dN}{dp d\epsilon} = \sigma A \frac{\sqrt{1-p^2/p_{\text{max}}^2}}{p}
    \label{eq:dipoleDist}
\end{equation}
for $p < p_\text{max}$ and zero otherwise; here, $\sigma$ is the TLS density with respect to energy and area $A$. Averaging over this distribution, we find a mean-square defect dipole moment $\expect{p^2} = p_{\rm max}^2/3$. 

For typical relative permittivities of aluminum oxide (9.8) and silicon dioxide (3.8), we expect the substrate-air loss to dominate the metal-air interface loss by an order of magnitude \cite{Martinis2022}. Moreover, we assume the dilute hydrofluoric acid etch performed immediately prior to metal deposition eliminates the disordered native oxide at the metal-substrate interface (see Appendix \ref{appendixA}). For this reason, we focus on loss from TLS defects at the substrate-air interface in the gap between qubit island and ground. The electric field in the gap $E(r)$ is well-approximated by that of a coplanar transmission line \cite{Murray2018}: 
\begin{equation}
     E(r) = \frac{V}{K'(r_i/r_o)} \frac{r_o}{\sqrt{|(r^2-r_i^2)(r^2-r_o^2)|}}.
    \label{eq:Efield}
\end{equation}
Here, $K'(r_i/r_o)$ is the complementary modulus to the elliptic integral of the first kind $K(k)$ such that $K'(k) = K(\sqrt{1-k^2})$. This expression agrees well with numerically calculated fields for our circular qubit geometry; see Appendix \ref{appendixC}. For separation from the gap edge $\left|r-r_{i,o}\right|~\ll~r_i,r_o$, we expect the electric field to scale as $E(r) \propto \left|r-r_{i,o}\right|^{-1/2}$, so that a cumulative histogram of $(2\pi/g)^2$ will be linear for small $(2\pi/g)^2$, corresponding to dominant TLS defects residing at the edges of the gap from qubit island to ground. We introduce the notation $\xi \equiv \left(2 \pi/g\right)^2$. We can show

\begin{align}
    \frac{dN}{d\xi \Delta f} &= \frac{2 \pi}{3h}  \, \sigma p_\text{max}^2 \, V_{\rm zp}^2 \, \frac{1}{K'(r_i/r_o)^2}\frac{r_o^2}{r_o^2-r_i^2},% \nonumber \\
    %&= \frac{\Gamma_1}{4 \pi^2 \alpha}.
    \label{eq:slope}
\end{align}
where $\Delta f$ is the frequency band of interest and $V_{\rm zp} = \sqrt{\hbar \omega_{01}/2C}$ is the zero-point voltage of the qubit mode with self-capacitance $C$. Similarly, in the continuum limit, a calculation from Fermi's Golden Rule yields a TLS contribution to the qubit relaxation rate given by 
\begin{equation}
    \Gamma_1 =  \frac{8\pi^3}{3 h} \, \sigma p_\text{max}^2 \, V_{\rm zp}^2\frac{\alpha }{K'(r_i/r_o)^2}\frac{r_o^2}{r_o^2-r_i^2},
\label{eq:fermi}
\end{equation}
where $\alpha$ is a geometric factor of order 10 that depends logarithmically on qubit geometry and the cutoff of the TLS distribution at the metal edges; see Appendix \ref{appendixD}. Comparison between Eqs.~\ref{eq:slope} and~\ref{eq:fermi} reveals a connection between the statistics of strongly-coupled TLS and the baseline relaxation rate away from dominant $T_1$ dropouts:
\begin{align}
    \frac{dN}{d\xi \Delta f} &= \frac{\Gamma_1}{4 \pi^2 \alpha}.
    \label{eq:eq6}
\end{align}

In Fig.~\ref{fig:f5}(a), we plot cumulative histograms of extracted $\xi \equiv (2\pi/g)^2$ for selected devices; the full dataset may be found in Appendix~\ref{appendixB}. Data from nominally identical qubits are combined, and the vertical axis is normalized to a 1~GHz band. As expected, we find shallower slopes in the distributions for larger-gap qubits. Additionally, we find that the distributions for the short-liftoff devices exhibit two distinct slopes, as seen especially clearly for the 100~$\upmu$m-gap devices. The steeper slope is well described by Eq.~\ref{eq:slope} with a defect density $\sigma$ = 2/GHz/$\upmu$m$^2$ and $p_\text{max}$ = 5~Debye (dashed line); previous experiments have found comparable values \cite{Martinis2005, Barends2013}. The shallower slope extending to lower $\left(2 \pi /g \right)^2$ (i.e., stronger coupling) represents a population of defects located very near the junction at the edges of the junction leads, where the zero-point electric fields are strongest. 

We perform Monte Carlo simulations of a transmon qubit coupled to a bath of TLS that reside at the SA interface and at the edges of the junction leads. We use Eq. \ref{eq:dipoleDist} to generate the proper distribution of qubit-TLS coupling for defects in the gap. To model defects at the junction leads, we use a linear density per unit energy $\lambda$ and the approximate field distribution at the edge of the leads \cite{Martinis2022}
\begin{equation}
    E(x) = \frac{1}{2} \frac{V}{\bar{r} \ln{4x/\bar{r}}},
    \label{eq:leadEfield}
\end{equation}
where $x$ is the distance from the junction and $\bar{r}$ is the width of the leads; see Appendix \ref{appendixC} for details. Figure \ref{fig:f5}(b) shows Monte Carlo simulation results plotted as a cumulative histogram of $(2 \pi/g)^2$ for the same geometries as in Fig. \ref{fig:f5}(a). For the 100~$\upmu$m-gap short lead devices, the filled circles include both gap and lead defects, while the blue ``+" symbols represent the lead defects alone. We assume defects in the gap are distributed with uniform density $\sigma=$~2/GHz/$\upmu$m$^2$ and we take values of $\lambda$ in the range 0.4-0.7/GHz/$\upmu$m to yield distributions that match the measured data. Our modeling shows that the dominant lead defects corresponding to $(2 \pi/g)^2<3\times10^8$~GHz$^{-2}$ are located within 500~nm of the junction. These results are supported by experiments from Lisenfeld, \textit{et al.}, which similarly conclude that dominant TLS are localized near the junctions \cite{Lisenfeld2025}. We believe that these defects are intrinsic to the additive liftoff process used to fabricate the qubit junctions. Moreover, these defects are responsible for the non-gaussian tails in our measured $T_1$ distributions. In Appendix~\ref{appendixC}, we present detailed modeling of the statistics of lead defects and their impact  on qubit $T_1$ distributions. 

\section{Summary and Outlook}
In summary, we have examined the dependence of energy relaxation time on qubit island and junction lead geometry. The devices described here involve aluminum base metal electrodes. However, since loss at the SA boundary dominates relaxation, we expect our results to apply to any superconducting qubits fabricated on silicon substrates, including those involving niobium and tantalum base metal layers. Strong coupling to individual TLS defects suppresses qubit $T_1$ at discrete operating frequencies; the TLS fluctuate on timescales of tens of minutes. For fixed qubit charging energy, devices with larger gap from qubit island to ground have longer $T_1$ times, consistent with a model of lossy defects that reside at interfaces. Furthermore, the $T_1$ distributions show low-$T_1$ tails that are dominated by defects residing at the edges of the leads within 500~nm of the junction. Continued improvements in qubit coherence will require new approaches to device fabrication that suppress the density of interfacial defects and avoid additive liftoff steps that are prone to leaving lossy organic residues.

\vspace{11pt}

\section{Acknowledgments}
This material is based upon work supported by the U.S. Department of Energy, Office of Science, National Quantum Information Science Research Centers and by the U.S. Department of Energy (DOE), Office of Science, Basic Energy Sciences under award no. DE-SC0020313. The authors gratefully acknowledge use of facilities and instrumentation in the UW-Madison Wisconsin Center for Nanoscale Technology (wcnt.wisc.edu). The Center is partially supported by the Wisconsin Materials Research Science and Engineering Center (NSF DMR-2309000) and the University of Wisconsin-Madison. This work made use of the Pritzker Nanofabrication Facility, part of the Pritzker School of Molecular Engineering at the University of Chicago, which receives support from Soft and Hybrid Nanotechnology Experimental (SHyNE) Resource (NSF ECCS-2025633), a node of the National Science Foundation’s National Nanotechnology Coordinated Infrastructure [RRID: SCR\_022955].

\section{Data Availability}

The data that support the findings of this study are openly available \cite{RawData}.

% \newpage

\clearpage
\appendix\section{ Device Fabrication}
\label{appendixA}

\begin{figure}[hb]
    \centering
    \includegraphics[width=\columnwidth]{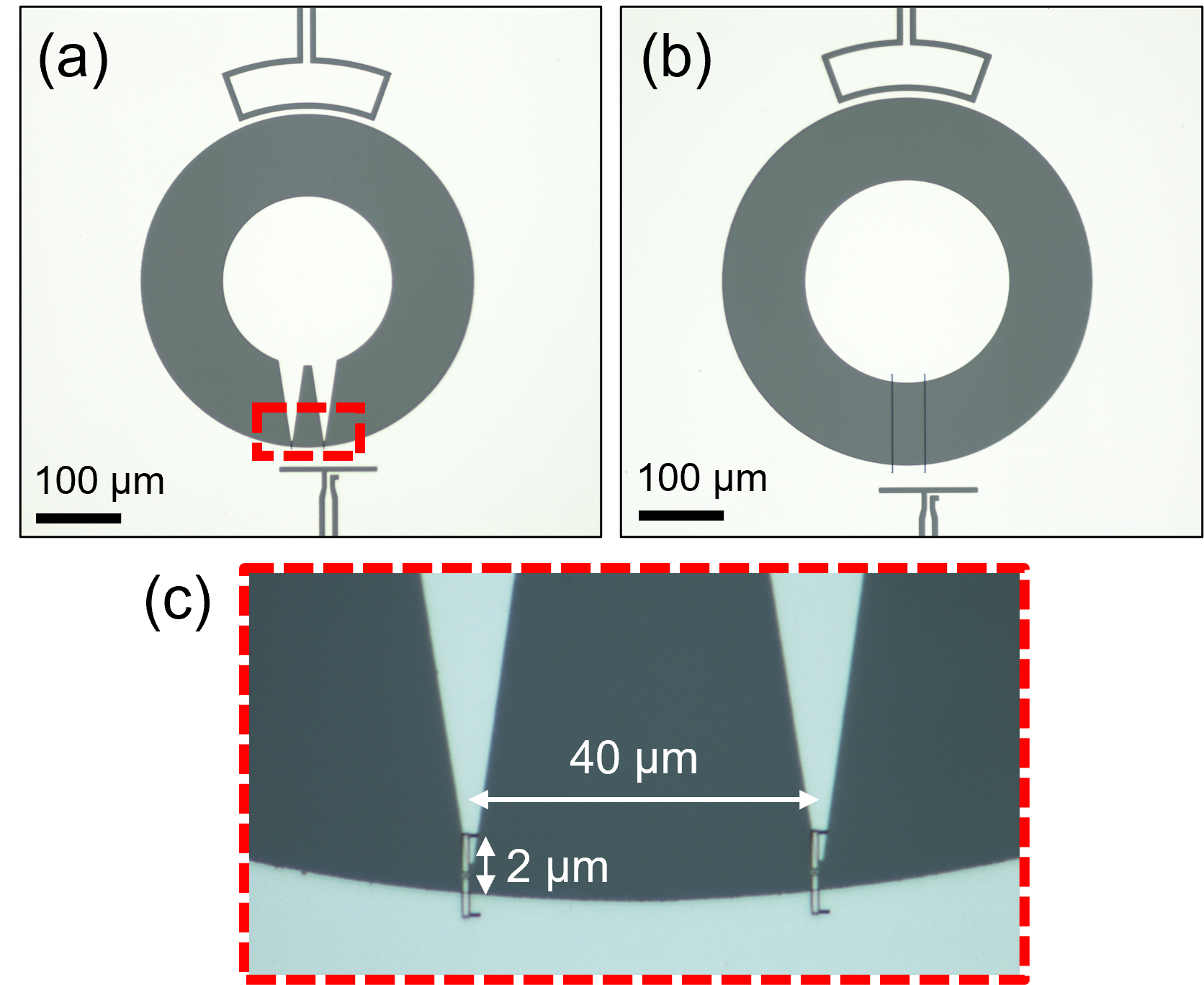}
    \caption{Optical micrographs of the 100~$\upmu$m-gap qubits. (a) 100~$\upmu$m-gap qubit with short-liftoff leads. (b) 100~$\upmu$m-gap qubit with long-liftoff leads. (c) Closeup view of the junctions for the short-liftoff qubit. The liftoff portion spans the 2~$\upmu$m gap from the groundplane to the tapered leads, which are defined in the same etch step as the qubit island.} 

    \label{fig:sup1}
\end{figure}

The qubits were fabricated on a 3-inch high-resistivity Si wafer ($>10$~k$\Omega$-cm). A 60~s etch in dilute (2\%) HF acid was used to remove the native silicon oxide immediately prior to sputter deposition of the 100-nm base aluminum layer. The qubit islands, bias leads, readout resonators, and feedline were defined using AZ-703 positive photoresist exposed on a Heidelberg DWL~66+ direct-write system. The wafer was developed and etched in the same step using AZ~300~MIF developer (2.38\% TMAH concentration). The Al-AlO$_x$-Al junctions were fabricated using a three-angle \textit{in-situ} bandaged Dolan bridge technique \cite{Bilmes2021} with a bilayer resist stack comprising MMA EL-13 and PMMA 950k A4. The resist was exposed on an Elionix 100~keV electron beam writer and developed for two minutes at 6$^\circ$C in 3:1 isopropanol:DI water. A descum ash and vapor HF acid etch were performed immediately prior to loading the wafer into the electron beam evaporator. The nominal junction areas are 0.04~$\upmu$m$^2$. Liftoff was performed in heated NMP followed by sonication in isopropanol. A final descum ash and vapor HF etch were performed after dicing and before packaging the dies. 
Figure \ref{fig:sup1}(a) shows an optical micrograph of the largest-gap qubit ($\Delta r$~=~100~$\upmu$m) with short liftoff leads. The area around the junctions is shown in Fig. \ref{fig:sup1}(b).

\section{Experimental Details}
\label{appendixB}

\begin{figure}
    \includegraphics[width=\columnwidth]{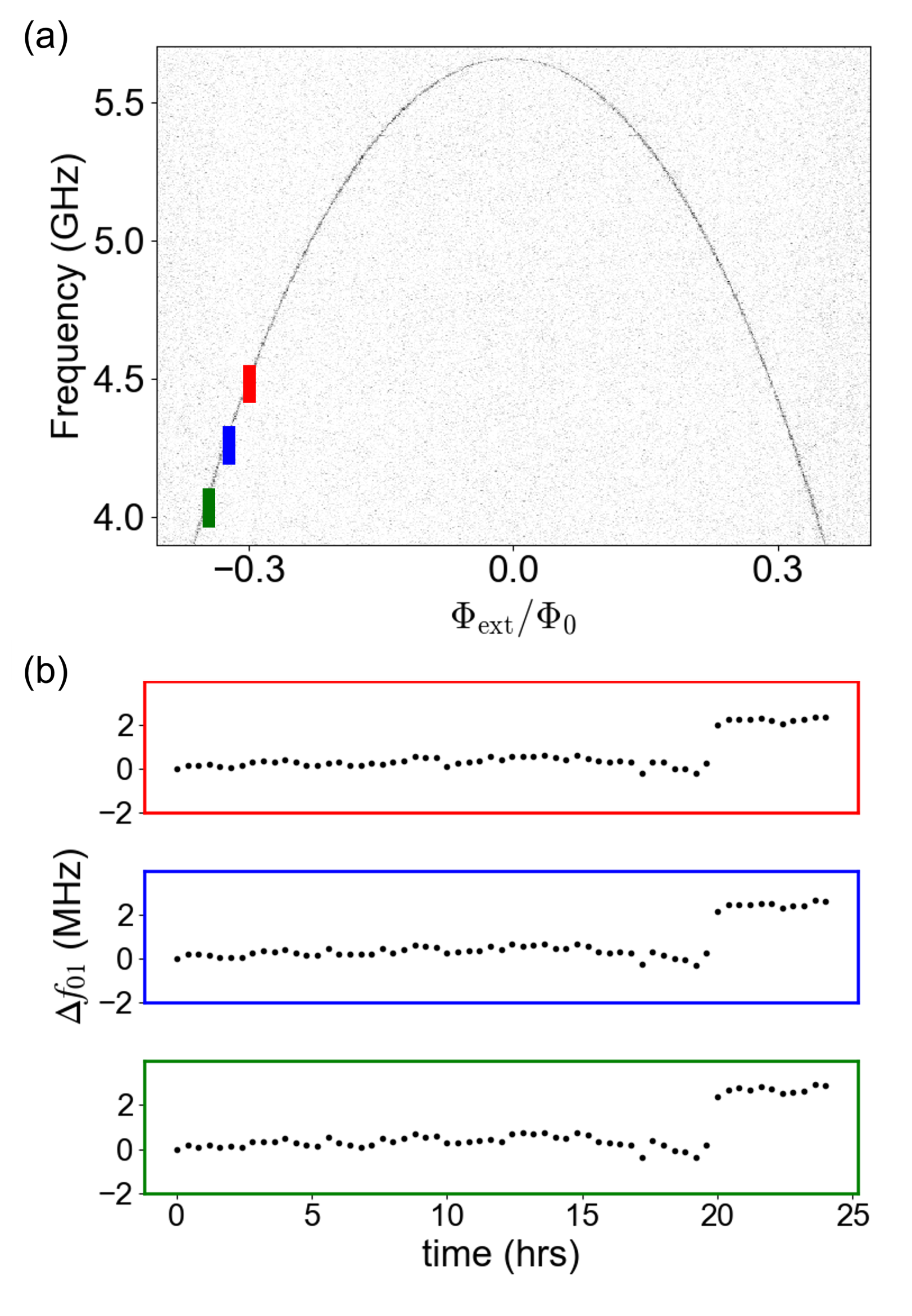}
    \caption{(a) Measured qubit flux dispersion, with colored lines indicating the three flux bias points at which repeated spectroscopy measurements were made in order to monitor and correct flux drift over long timescales. (b) Time series of extracted qubit frequency shifts at the three bias points corresponding to the colored regions in (a). Each dot is obtained from a Lorentzian fit to the qubit spectrum.} 
    \label{fig:sup2}
\end{figure}

The data shown in Fig. \ref{fig:f3} represent 24 hours of repeated $T_1$ swap scans. Qubit idle points are chosen to yield 1~MHz frequency resolution over a 500~MHz span; at each idle point, 40 separate inversion-recovery experiments are conducted with 50 averages per idle delay. Multiple qubits on the same chip are measured in an interleaved fashion. A single swap scan to extract $T_1$ time \textit{versus} frequency for all qubits takes around 20 minutes to complete. Prior to each swap scan, we perform qubit spectroscopy at multiple bias points for each device in order to correct for any long-term flux drifts that would cause correlated shifts in qubit frequency that are distinct from those due to the spectral diffusion of TLS defects. Figure \ref{fig:sup2}(a) shows an example qubit spectrum with colored boxes indicating the bias regions used to monitor long-term flux drift. Figure \ref{fig:sup2}(b) shows the frequency shifts measured on one qubit at three different operating points over a period of 24 hours. This dataset is chosen because of the jump in qubit frequency around hour 20. This jump was observed only in this particular qubit; moreover, the jump magnitude scales linearly with the qubit flux-to-frequency transfer function $df_{01}/d\Phi$ at the three different bias points. We thus conclude that the frequency jump is due to a local flux offset in the compound Josephson junction of the qubit device, as opposed to a sudden TLS reconfiguration as might occur, e.g., following a gamma-ray impact \cite{Thorbeck2023}. For qubit bias at an operating frequency of 4.25~GHz (center trace, blue bounding box), the flux-to-frequency transfer function is around 5~GHz/$\Phi_0$, so that frequency fluctuations of 1~MHz correspond to flux fluctuations of order 200~$\upmu \Phi_0$. 

\begin{figure}
    \centering
    \includegraphics[width=\columnwidth]{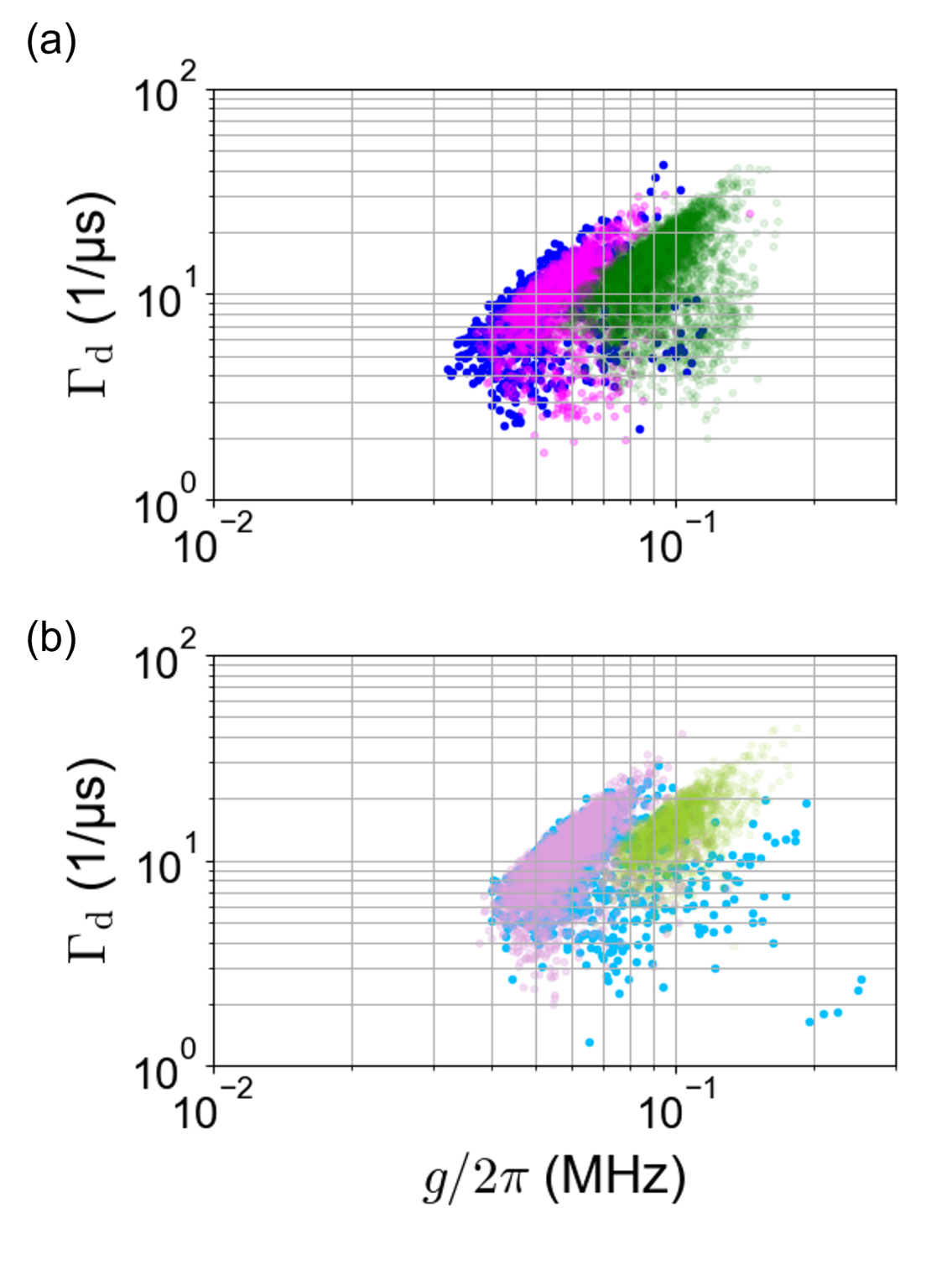}
    \caption{Extracted defect lifetime $\Gamma_\text{d}$ \textit{vs.} coupling $g/2\pi$. (a) Short-liftoff qubits with $\Delta r$ = 5 $\upmu$m (green), 20 $\upmu$m (pink), and 100 $\upmu$m (blue). (b) Long-liftoff qubits with $\Delta r$ = 5 $\upmu$m (green), 20 $\upmu$m (pink), and 100 $\upmu$m (blue).}
    \label{fig:supf3}
\end{figure}

\begin{figure}
    \centering
    \includegraphics[width=\columnwidth]{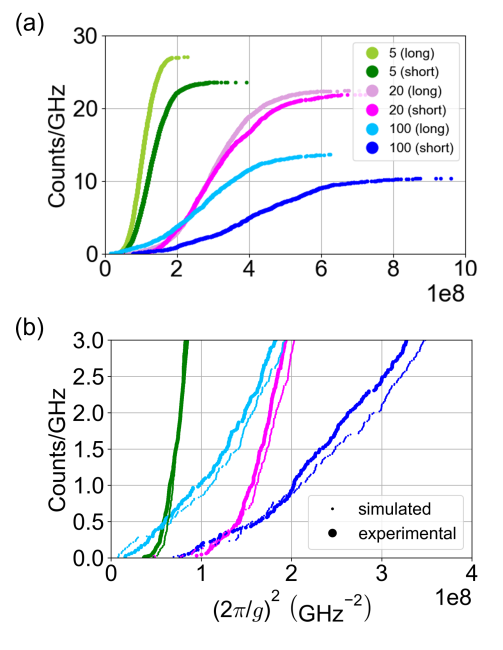}
    \caption{(a) Cumulative histograms of $(2\pi/g)^2$ for all measured qubits, with data from nominally identical qubits combined into a single trace. The legend indicates gap $\Delta r$ from qubit island to ground in microns, with qubit lead geometry shown in parentheses. (b) Comparison of measured data and Monte Carlo simulations at low $(2\pi/g)^2$; this plot combines the data from Figs.~\ref{fig:f5}(a-b) on a single set of axes.}
    \label{fig:g2dist}
\end{figure}

We extract the parameters $g$ and $\Gamma_\text{d}$ of strongly-coupled TLS by fitting prominent peaks in the frequency sweeps of $\Gamma_1$,   with a threshold at one standard deviation from the mean; example fits are shown in the inset of Fig. \ref{fig:f2}(b). We plot extracted $g/2\pi$ and $\Gamma_\text{d}$ for all qubits in Figs.~\ref{fig:supf3}(a) (short-liftoff qubits) and \ref{fig:supf3}(b) (long-liftoff qubits). The average TLS-qubit coupling strength decreases with increasing gap, as expected. For the long-liftoff qubits with 100 $\upmu$m gap, we see a population of high-$g$ TLS that we attribute to strongly coupled defects near the junction leads. The extracted couplings are plotted in Figs.~\ref{fig:f5}(a) and \ref{fig:g2dist} as integrated histograms of $(2\pi/g)^2$. The defect counts for a given qubit geometry are calculated by collecting extracted $g$ values from fits to every scan of $\Gamma_1$ \textit{versus} frequency, dividing the counts by the number of time traces, and normalizing to a 1~GHz band.

\section{Details of Monte Carlo Modeling}
\label{appendixC}

We perform Monte Carlo simulations to model qubit-TLS interactions for two populations of defects: those that live at the SA interface in the qubit gap and those at the edges of the Josephson junction leads. The dipole moment $p$ of each defect is generated from the distribution of Eq. \ref{eq:dipoleDist}. Gap defects are randomly placed at the SA interface assuming a uniform surface density and at the edges of the junction leads with uniform linear density. 

\subsection{Calculation of Electric Fields}

\begin{figure}
    \centering
    \includegraphics[width=\columnwidth]{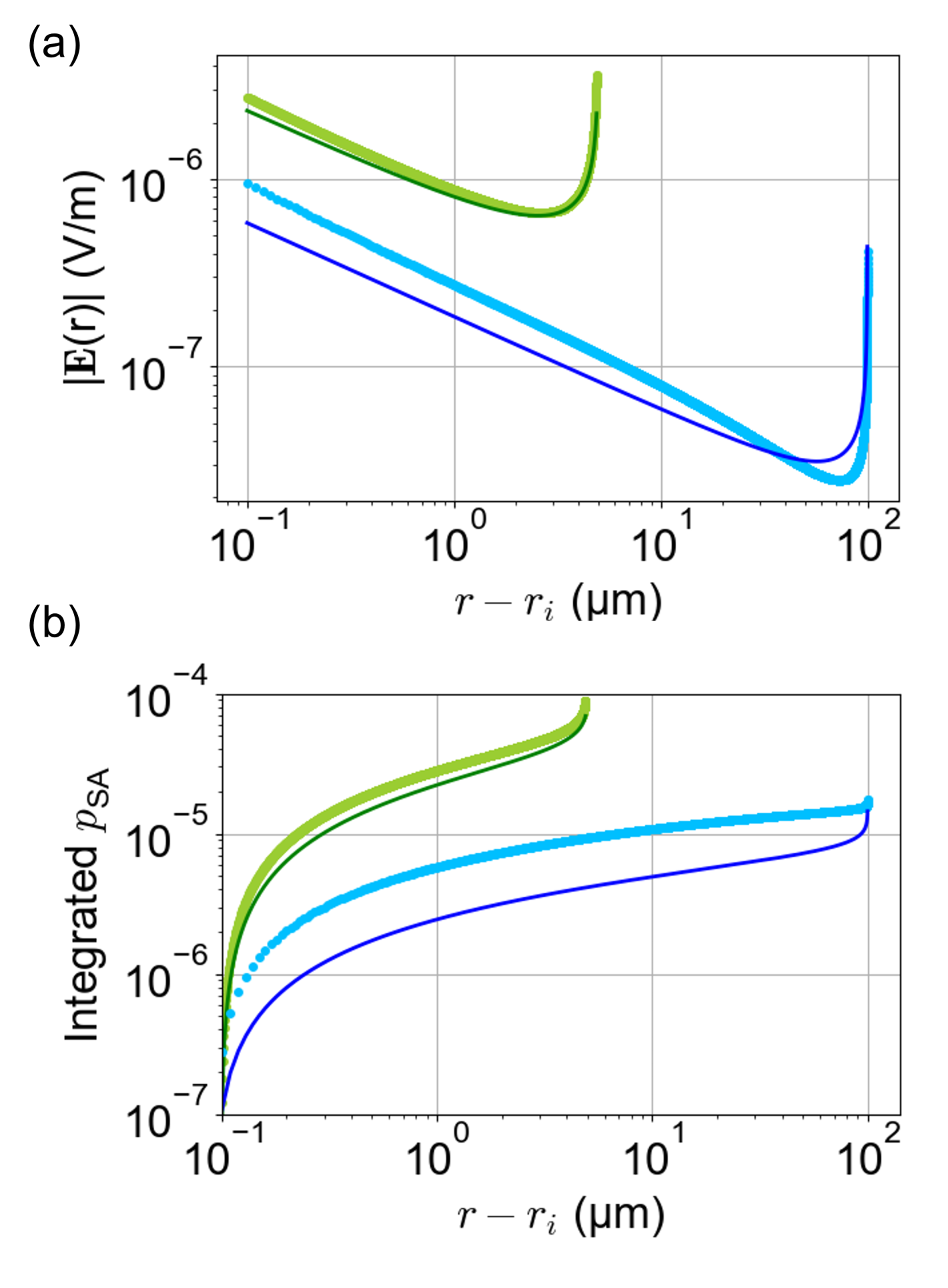}
    \caption{(a) Electric field magnitude in the gap as a function of distance from the island $r - r_i$ for qubit gaps of 5~$\upmu$m (green) and 100~$\upmu$m (blue). The closed circles are calculated numerically, while the lines are generated using Eq. \ref{eq:Efield}. The absolute field scale is set by the zero-point voltage on the qubit island $V_{\rm zp} = \sqrt{\hbar \omega_{01}/2C}$, where we take $\omega_{01}/2\pi =$~4.25~GHz and $C=$~70~fF. (b) Integrated SA participation ratio as a function of distance from the island $r - r_i$ for qubit gaps of 5 $\upmu$m and 100 $\upmu$m.}
    \label{fig:Ecomparison}
\end{figure}

For our modeling, we use the analytical expressions of Eqs.~\ref{eq:Efield} and \ref{eq:leadEfield} to model electric fields in the gap from qubit island to ground and at the edges of the junction leads, respectively. In Fig.~\ref{fig:Ecomparison}, we compare electric field strength calculated from the coplanar expression Eq.~\ref{eq:Efield} to that calculated numerically; we find that the integrated participation ratio for these two cases differs by 23\% for $\Delta r$~= 5~$\upmu$m and by 17\% for $\Delta r$~= 100~$\upmu$m. In the next subsection, we compare the model of edge defects near the junction to a model involving an enhanced surface density of defects near the junction leads.

\subsection{Modeling of Lead Defects}
\label{2}

In our Monte Carlo simulations, we assume defects that are distributed uniformly along the edges of the junction leads. We use the electric field distribution of Eq.~\ref{eq:leadEfield}, where we take $\bar{r}=0.3$ $\upmu$m to match the width of the junction leads for these devices. For the short-liftoff qubits, we take the lead length to be 2~$\upmu$m, which is the distance from the optically-defined leads to the ground plane. For the long-liftoff qubits, we take the lead length to be $\Delta r$. We find that $\lambda$ in the range from 0.4-0.7/GHz/$\upmu$m provides the best match to the experimental data. 

\begin{figure}
    \centering
    \includegraphics[width=\columnwidth]{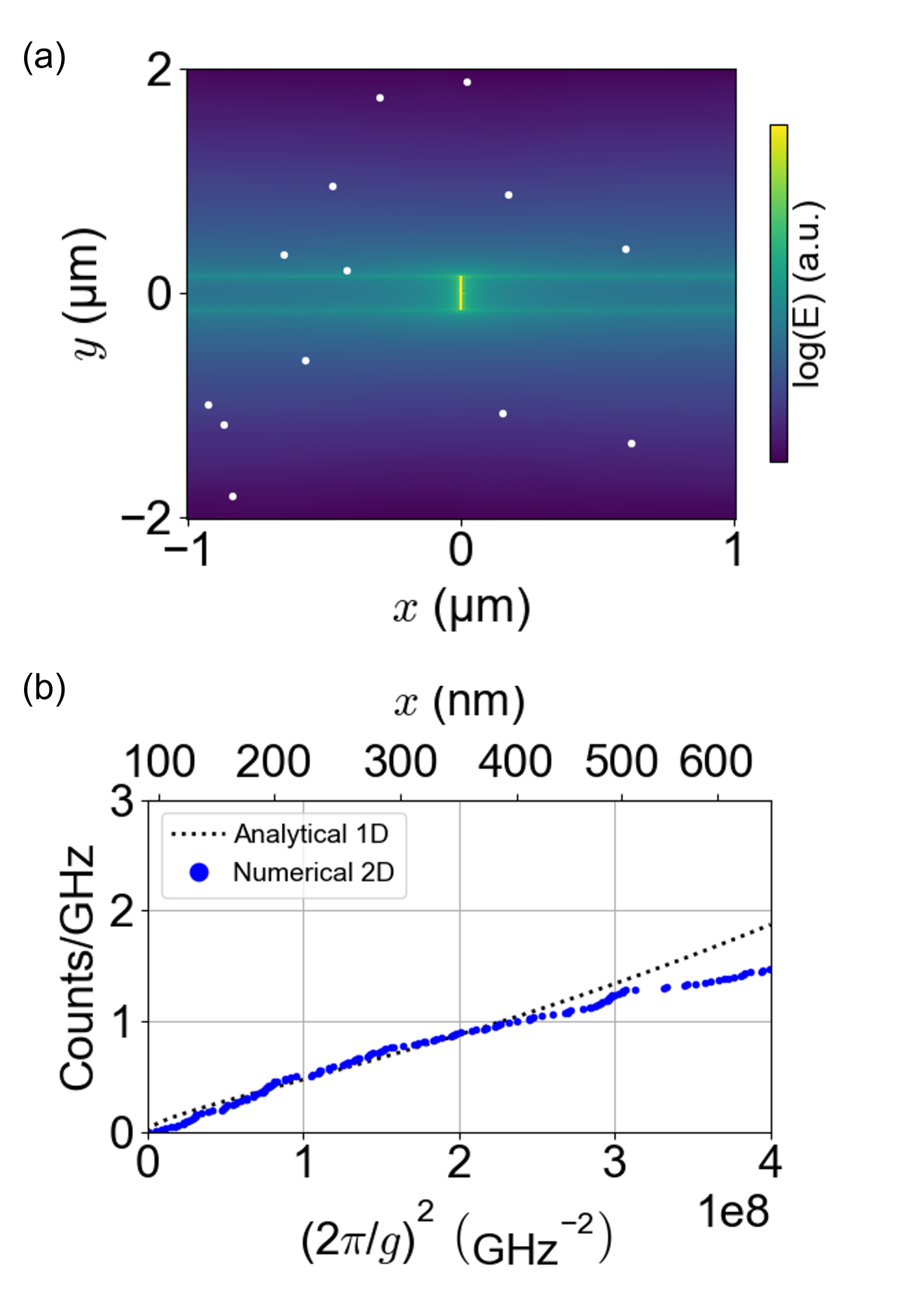}
    \caption{(a) Numerically simulated electric fields around the junction leads, which are oriented along the $x$-direction with the junction at $x=0$. The white dots represent TLS defects placed randomly near the leads (here with areal density 3.2~$\upmu$m$^{-2}$). (b) Cumulative distribution of $(2\pi/g)^2$ generated using an analytical expression for 1D edge defects (dotted line) and for a 2D model of surface defects (blue symbols). For the 1D model, we take $\lambda$ = 0.4/GHz/$\upmu$m, while for the 2D model we take $\sigma$ = 3.2/GHz/$\upmu$m$^2$.}
    \label{fig:ansysLeads}
\end{figure}

To compare the model of defects distributed along the edges of the junction leads to a model of defects distributed uniformly in the area around the junctions, we simulate the distributions of inverse-square coupling $\left(2 \pi/g\right)^2$ for these two cases. For the linear model, we take the $E$-field distribution of Eq.~\ref{eq:leadEfield}; for the area model, we use Ansys Maxwell 3D to calculate the electric fields for leads with width~0.3~$\upmu$m, thickness 0.1~$\upmu$m, and length~2~$\upmu$m. Figure \ref{fig:ansysLeads}(a) shows the electric fields with a sample distribution of defects represented as white dots. A surface density of defects of 3.2/GHz/$\upmu$m$^2$ produces a distribution of $(2\pi/g)^2$ that is well matched to a linear edge density $\lambda = 0.4$/GHz/$\upmu$m, as shown in Fig. \ref{fig:ansysLeads}(b).  We find that the tails in our defect distributions across all devices are fit by linear edge density in the range $\lambda = 0.4-0.7$/GHz/$\upmu$m, corresponding to surface density of defects in the range 3-6/GHz/$\upmu$m$^2$ in the immediate vicinity of the junctions, significantly higher than the defect density across most of the qubit gap. The higher defect density in the junction area points to liftoff residues as a likely source; the variation in the inferred edge density of defects is likely due to intrinsically poor control of the metal liftoff.

\subsection{Check on Extraction of Qubit-TLS Coupling}

In our simulations, TLS defects are placed randomly in the gap from qubit island to ground or along the qubit leads; coupling strength $g$ is calculated for each defect; and the defect is assigned a frequency from a uniform distribution. Defect lifetimes $\Gamma_\text{d}$ are drawn from a Gaussian distribution centered at 5~$\upmu$s$^{-1}$ to match the experimental data. We take a mean-square dipole moment of $p_\text{max}^2/3$ with $p_\text{max}$ = 5 Debye; see Appendix C.4 below. In Fig. \ref{fig:f7}(a), we show an example $\Gamma_1$ trace from one simulation run. We fit a Lorentzian lineshape to peaks in $\Gamma_1$ that are one standard deviation above the mean; the fits are shown in red. In Fig. \ref{fig:f7}(b), the generated values of coupling strength are compared to the values extracted from fits to the generated data. This comparison acts as a check to see how faithfully our fitting code can extract defect parameters. The extracted coupling strengths match the generated values to within a few percent, with a small bias toward higher extracted $g$. This gives us confidence that the fits to our experimental data are reasonable.

\begin{figure}
    \includegraphics[width=\columnwidth]{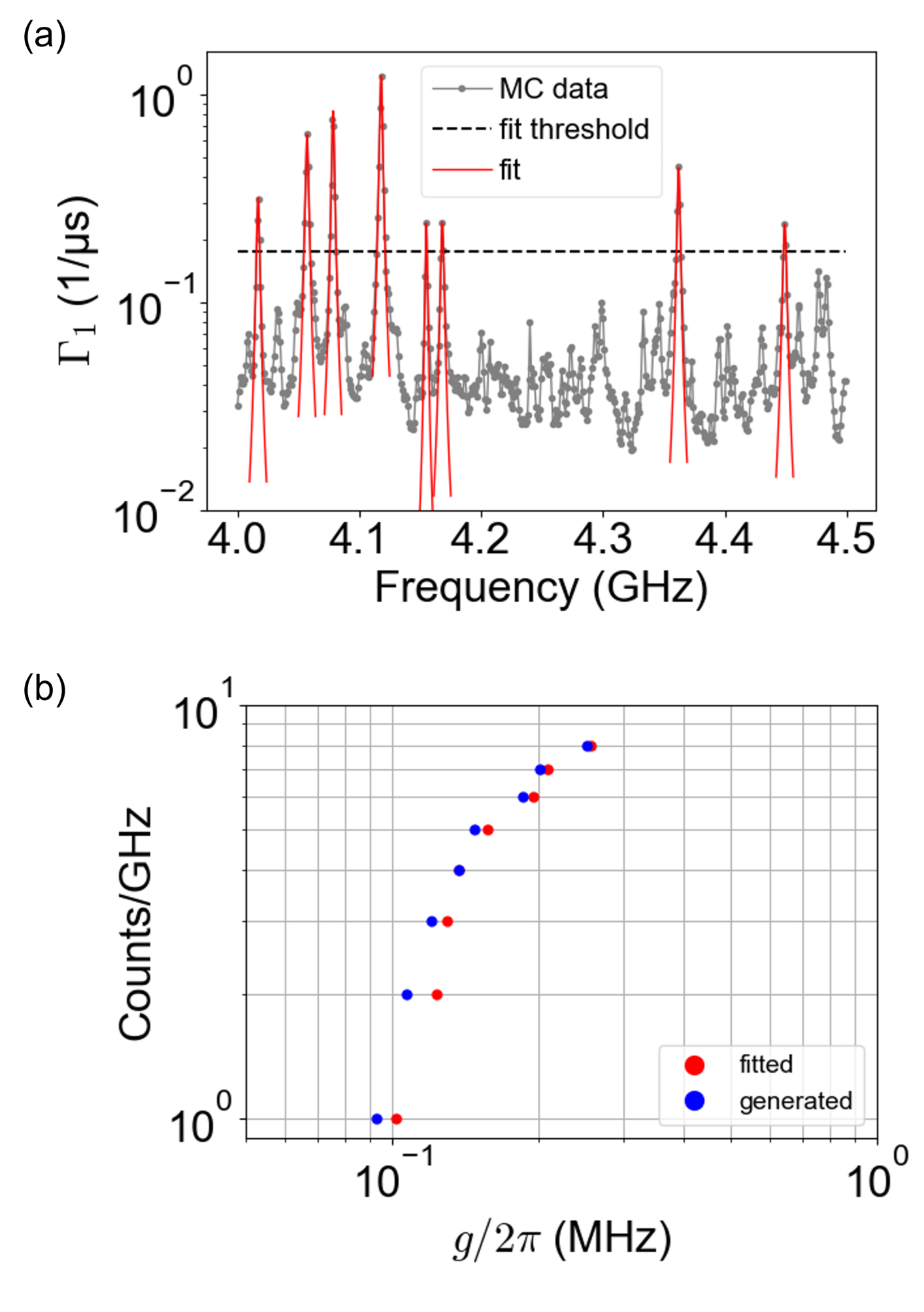}
    \caption{(a) Simulated qubit relaxation rate $\Gamma_1$ from Monte Carlo modeling. Here, peaks exceeding 1 standard deviation from the mean are automatically fit with a Lorentzian lineshape to extract coupling $g$ and defect lifetime $\Gamma_\text{d}$. The fits are shown in red. (b) Check on extraction of coupling strength $g$. Blue symbols are generated by Monte Carlo modeling. Red symbols are obtained from fits to the generated data. In most cases, the fit value differs from the generated value by a few percent.}
    \label{fig:f7}
\end{figure}

\subsection{Separate Determination of $p_{\rm max}$ and $\sigma$}

\begin{figure}
    \centering
    \includegraphics[width=\columnwidth]{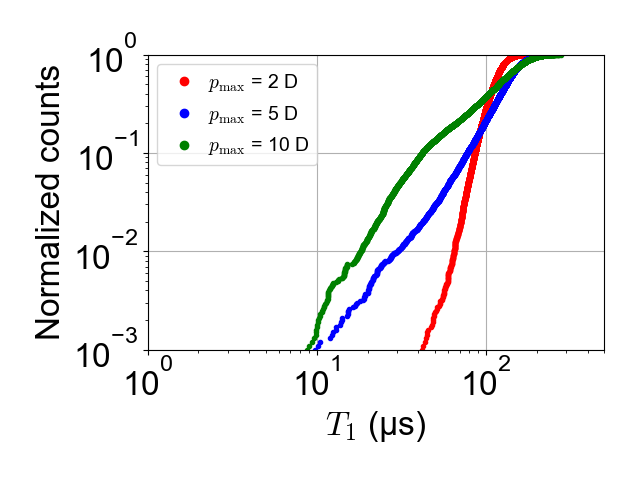}
    \caption{Simulated cumulative $T_1$ distributions for the 100-$\upmu$m gap short-liftoff geometry. Here we fix the product of $p_\text{max}^2$ and $\sigma$, and vary $p_\text{max}$. The number of defects sets the distribution width. The simulated and measured data match well for $p_\text{max}$ = 5~Debye, $\sigma = $ 2/GHz/$\upmu$m$^2$.}
    \label{fig:t1dist_MC}
\end{figure}

\begin{figure}
    \centering
    \includegraphics[width=\columnwidth]{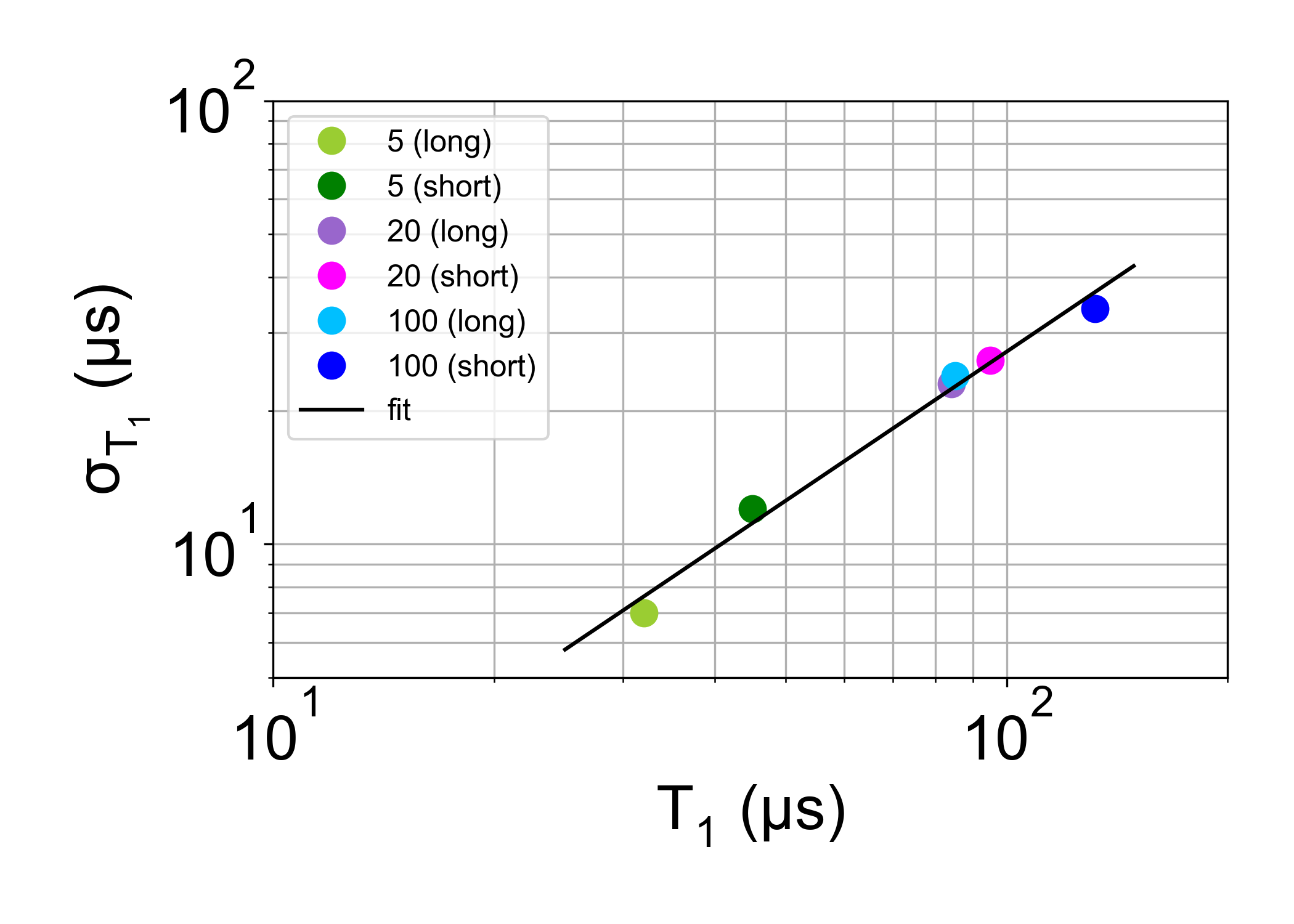}
    \caption{Standard deviation in $T_1$ \textit{vs.} mean $T_1$ for all six qubit geometries, where data from nominally identical qubits have been combined. The data is fitted with a power law with an exponent of 1.1.}
    \label{fig:stdev_vs_t1}
\end{figure}

Qubit mean $\Gamma_1$ and the slopes in the distributions of $\left(2 \pi/g\right)^2$ are both proportional to $\sigma p_{\rm max}^2$; however, the width of the measured $T_1$ distributions sets another constraint that provides information about $\sigma$ and $p_{\rm max}$ separately. We perform Monte Carlo simulations of the 100~$\upmu$m-gap short-liftoff geometry, fixing $\sigma p_\text{max}^2$ and varying $p_\text{max}$. Figure \ref{fig:t1dist_MC} shows simulated $T_1$ distributions for three values of $p_\text{max}$; for $p_\text{max} = 5$ Debye and $\sigma = $ 2/GHz/$\upmu$m$^2$, the standard deviation is 37 $\upmu$s, which is close to the experimental standard deviation of 34 $\upmu$s. For $p_\text{max}$ = 2 Debye and 10 Debye, the standard deviations are 20 $\upmu$s and 52 $\upmu$s, respectively.
Figure \ref{fig:stdev_vs_t1} shows the measured standard deviation in $T_1$ \textit{versus} mean $T_1$ for all six qubit geometries.  For a fixed qubit geometry with varying TLS density, one can show that $\sigma_{T_1} \propto T_1^{3/2}$ \cite{You2022, Bal2024, YuCarruzco2025}. In contrast, here we vary qubit geometry while maintaining a fixed defect density. For a larger gap $\Delta r$ from qubit island to ground there are more defects; however, qubit $T_1$ increases due to weaker qubit-TLS coupling. Thus we expect a dependence of $\sigma_{T_1}$ on $T_1$ that is somewhat weaker than the 3/2 power. Indeed, we find qubit $T_1$ spread shows a power law dependence on $T_1$, namely $\sigma_{T_1} \propto T_1^{1.1}$. 

\subsection{Contribution of Lead Defects to $T_1$ Tails}

\begin{figure*}
    \centering
    \includegraphics[width=0.95\linewidth]{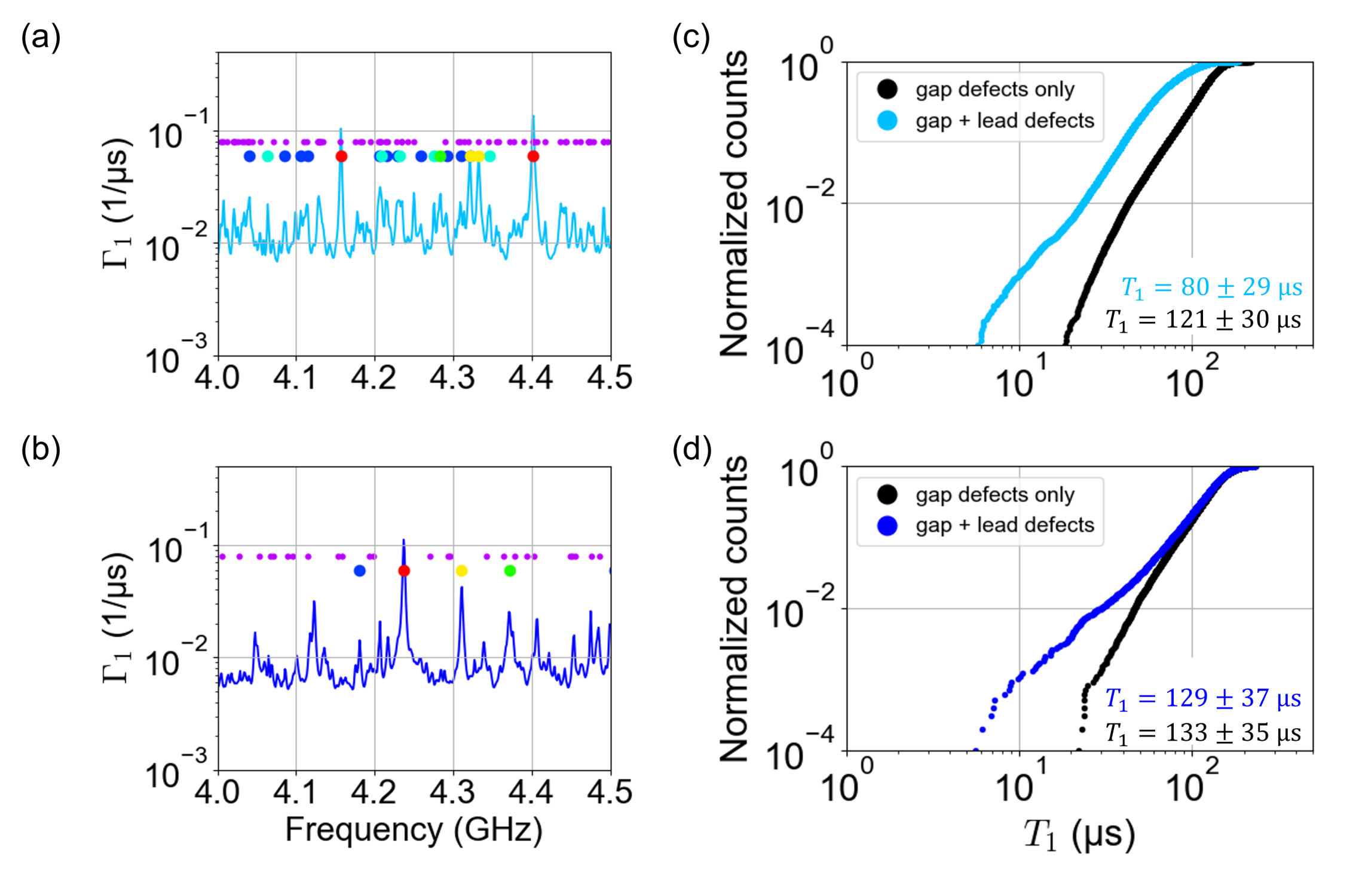}
    \caption{Monte Carlo simulations of $\Gamma_1$ \textit{versus} qubit operating frequency for qubits with gap $\Delta r$ = 100~$\upmu$m and (a) long-liftoff leads or (b) short-liftoff leads. The colored symbols indicate the frequencies of lead defects, where color denotes distance $x$ from the junction: red for $x < 0.5$ $\upmu$m; yellow for $0.5 \leq x < 1$ $\upmu$m; green for $1 \leq x < 2$ $\upmu$m; cyan for $2 \leq x < 5$ $\upmu$m; blue for $5 \leq x < 10$ $\upmu$m; and purple for $x \geq 10$ $\upmu$m. (c,d) Simulated $T_1$ distributions for (long-liftoff, short-liftoff) qubits, respectively. The black traces are generated from a bath of gap defects alone, while the colored traces take into account both gap and lead defects. The addition of lead defects results in nongaussian tails in the $T_1$ distributions.}
    \label{fig:t1dist_leadsNoleads}
\end{figure*}

In Figure~\ref{fig:t1dist_leadsNoleads}, we plot simulated $\Gamma_1$ \textit{versus} operating frequency for (a) the 100~$\upmu$m-gap long-liftoff geometry and (b) the 100~$\upmu$m-gap short-liftoff geometry. Both lead and gap defects are considered. For both devices, we take gap defect density 2/GHz/$\upmu$m$^2$. For the short-liftoff qubit, we take lead defects with density 0.4/GHz/$\upmu$m for the 2-$\upmu$m liftoff portion of the leads and we take defect density 0.25/GHz/$\upmu$m for the optically-defined tapered portion of the leads, corresponding to the surface density 2/GHz/$\upmu$m$^2$ (see Appendix \ref{appendixC}.2 above). For the tapered leads, we again use Eq.~\ref{eq:leadEfield} to calculate the edge fields, but we take the effective trace width $\bar{r}$ to be dependent on distance from the junction to account for the taper. For the long-liftoff qubit, we take lead defect density 0.6/GHz/$\upmu$m. Overlaid on the plot of $\Gamma_1$, we indicate the frequencies of lead defects using colored symbols; see figure caption for details. Figure~\ref{fig:t1dist_leadsNoleads} (c,d) shows simulated $T_1$ distributions for the same devices. These plots were generated by averaging 30 separate realizations of defect disorder. Here, the black traces take into account defects in the gap alone, while the colored traces include both gap and lead defects. The distributions generated from gap defects alone are close to Gaussian; however, when lead defects are added, the distributions shift to shorter $T_1$ with long tails extending to $T_1$ of order a few $\upmu$s. The lead defects that dominate $T_1$ dropouts are located within 500~nm of the junction. 

\section{Analytic Expression for $\Gamma_1$ due to SA Defects}
\label{appendixD}

The qubit relaxation rate $\Gamma_1$ can be expressed using Fermi's golden rule as

\begin{equation}
    \Gamma_1 = \frac{4 \pi^2 \sigma}{\hbar} \int_{r_i}^{r_o}{\braket{p^2} E^2(r) \, r dr}
\end{equation}
where we use the mean-square dipole moment $\braket{p^2}$ = $\int_0^{p_\text{max}}p^2f(p)dp = p_\text{max}^2/3$, where $f(p)$ is given by Eq. \ref{eq:dipoleDist}. We take $E(r)$ given by Eq. \ref{eq:Efield} and introduce a small cutoff $\delta$ to avoid the logarithmic divergence at the metal edges, yielding
\begin{equation}
    \Gamma_1 = \frac{8\pi^3}{3h} \sigma \,p_\text{max}^2 V_{\rm zp}^2 \, \frac{r_o^2}{K'(r_i/r_o)^2} \int_{r_i+\delta}^{r_o-\delta}{\frac{rdr}{(r^2-r_i^2)(r_o^2-r^2)}}.
\end{equation}
Performing the integral, we find

\begin{equation}
    \Gamma_1 =  \frac{8\pi^3}{3 h} \, \sigma p_\text{max}^2 \, V_{\rm zp}^2\frac{\alpha }{K'(r_i/r_o)^2}\frac{r_o^2}{r_o^2-r_i^2},
\label{eq:supFermi}
\end{equation}
with $\alpha$ = $\log{\frac{r_0^2-r_i^2}{\delta \sqrt{2r_ir_o}}}$.

We use this expression to calculate qubit $T_1$ as a function of gap $\Delta r$ in Fig. \ref{fig:Figure1}(d). For gap size $\Delta r$ ranging from 1 $\upmu$m to 300 $\upmu$m, we choose appropriate $r_i$ and $r_o$ values to maintain a constant charging energy $E_c/h =$ 270~MHz. To determine these values, we note that the capacitance $C = e^2/2E_c$ of a circular island in a ground plane may be calculated by mapping from the self-inductance of the dual structure, a flat circular washer \cite{Rafferty2021}. Analytic expressions for the self-inductance of a flat circular washer are worked out in the literature, e.g. by Brojeny and Clem \cite{BrojenyClem2003}. We can thus create a lookup table to map from $\Delta r$ to $r_i, r_o$ and then from Eq.~\ref{eq:supFermi} to $\Gamma_1$. 

\section{Cumulative distribution of $\left(2 \pi/g\right)^2$ for lead defects}
\label{appendixE}

We consider defects uniformly distributed along the junction leads with density $\lambda$ per unit energy per unit length. Similarly, we use Eq.~\ref{eq:leadEfield} to model electric fields $E(x)$ at the edges of the leads a distance $x$ from the junction, and we take the average defect dipole moment $\expect{p^2} = p_{\rm max}^2/3$. We introduce the notation $\xi \equiv \left(2\pi/g\right)^2$. We have

\begin{equation}
    \frac{dN}{d \xi} = h \Delta f \, \lambda \, \frac{dx}{d \xi},
\end{equation}
with
\begin{align}
    \xi (x) &= \frac{3 h^2}{p_{\rm max}^2E^2(x)} \nonumber \\
    &= 12h^2 \frac{\bar{r}^2}{p_{\rm max}^2V_{\rm zp}^2}\ln^2\left(\frac{4 x}{\bar{r}}\right).
\end{align}

Integrating over the leads, we find 
\begin{equation}
    N\left(\xi<\xi'\right) = \frac{h \Delta f}{4} \lambda \bar{r} \left[ \exp\left(\frac{V_{\rm zp} p_{\rm max}}{2\sqrt{3} h \bar{r}} \sqrt{\xi'}\right) - 1\right].
\end{equation}
This expression is used to generate the dotted trace in Fig.~\ref{fig:ansysLeads}(b).

\section{TLS Spectral Diffusion}
\label{appendixF}

\begin{figure}
    \centering
    \includegraphics[width=\columnwidth]{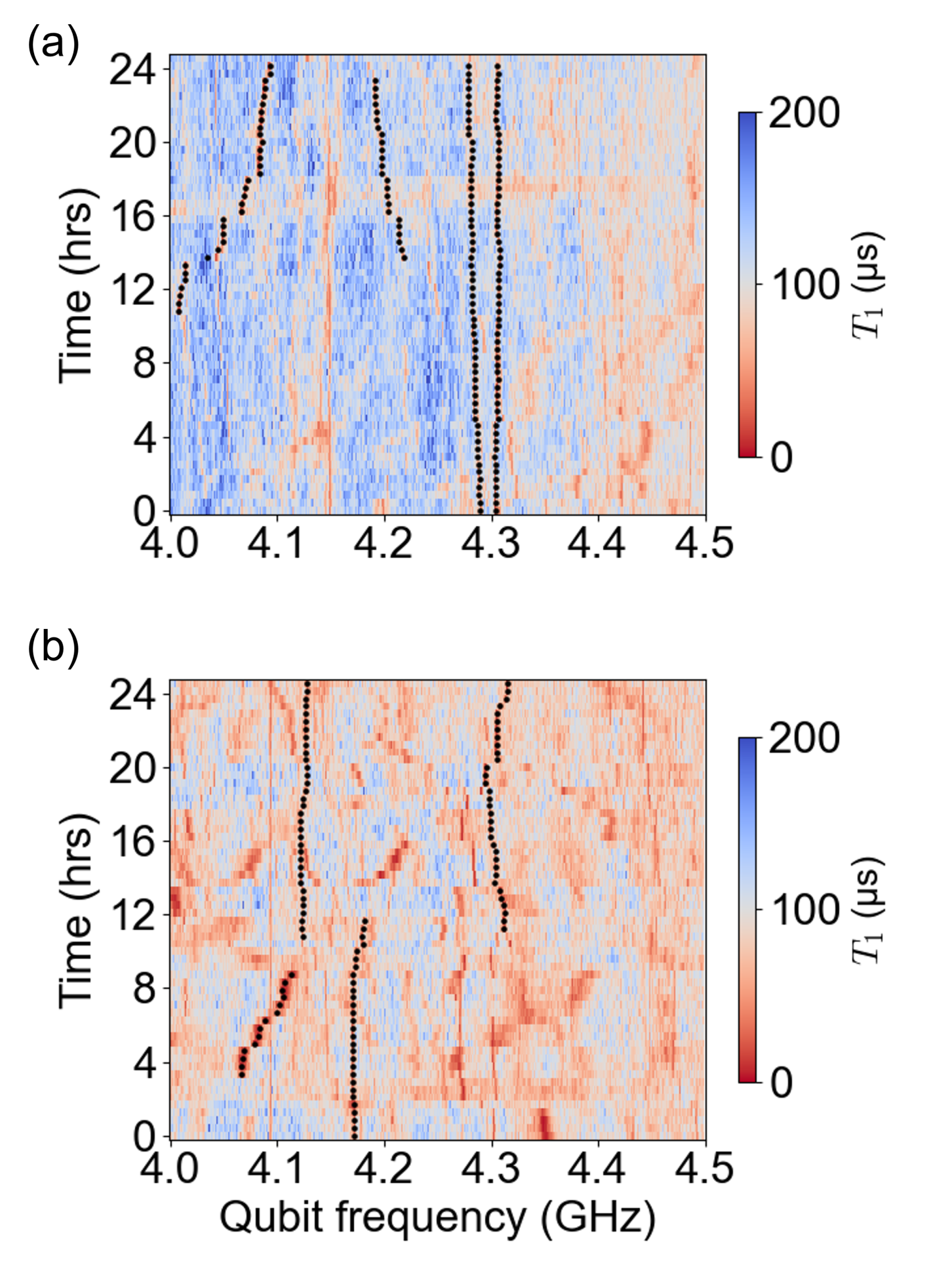}
    \caption{Repeated swap spectroscopy scans for qubits with $\Delta r = 100$ $\upmu$m and (a) short-liftoff leads and (b) long-liftoff leads. Four defects are chosen from each dataset and their resonant frequencies (black dots) are tracked over time to extract TLS diffusivity.}
    \label{fig:diffusion}
\end{figure}

We track the resonant frequency of eight defects over time to characterize TLS spectral diffusion. Figure~\ref{fig:diffusion} shows repeated swap data from two qubits with $\Delta r$ = 100~$\upmu$m and (a) short-liftoff leads and (b) long-liftoff leads. For each scan, we track the resonant frequency of several defects (black dots). In general, the TLS dynamics are diffusive, consistent with coupling to a bath of TF. In some cases, TLS diffusion is accompanied by a large linear drift, which we speculate could be due to strain relaxation in the sample. After subtracting out any linear drift, we extract TLS diffusivity from a fit to the relation $\sigma_{f_\mathrm{d}}(t) = 2D\sqrt{t}$. An average over eight defects yields a diffusivity $D$ = 2.2 $\pm$ 0.1 MHz/hr$^{1/2}$. This agrees well with the diffusivity $D$ = 2.5 $\pm$ 0.1 MHz/hr$^{1/2}$ reported by Klimov \textit{et al.} \cite{Klimov2018}.

\clearpage

\bibliographystyle{naturemag_local}
\bibliography{references}

@article{Martinis2005,
  title = {Decoherence in Josephson Qubits from Dielectric Loss},
  author = {Martinis, John M. and Cooper, K. B. and McDermott, R. and Steffen, Matthias and Ansmann, Markus and Osborn, K. D. and Cicak, K. and Oh, Seongshik and Pappas, D. P. and Simmonds, R. W. and Yu, Clare C.},
  journal = {Phys. Rev. Lett.},
  volume = {95},
  issue = {21},
  pages = {210503},
  numpages = {4},
  year = {2005},
  month = {Nov},
  publisher = {American Physical Society},
  doi = {10.1103/PhysRevLett.95.210503},
  url = {https://link.aps.org/doi/10.1103/PhysRevLett.95.210503}
}

@article{Klimov2018,
  title = {Fluctuations of Energy-Relaxation Times in Superconducting Qubits},
  author = {Klimov, P. V. and Kelly, J. and Chen, Z. and Neeley, M. and Megrant, A. and Burkett, B. and Barends, R. and Arya, K. and Chiaro, B. and Chen, Yu and Dunsworth, A. and Fowler, A. and Foxen, B. and Gidney, C. and Giustina, M. and Graff, R. and Huang, T. and Jeffrey, E. and Lucero, Erik and Mutus, J. Y. and Naaman, O. and Neill, C. and Quintana, C. and Roushan, P. and Sank, Daniel and Vainsencher, A. and Wenner, J. and White, T. C. and Boixo, S. and Babbush, R. and Smelyanskiy, V. N. and Neven, H. and Martinis, John M.},
  journal = {Phys. Rev. Lett.},
  volume = {121},
  issue = {9},
  pages = {090502},
  numpages = {5},
  year = {2018},
  month = {Aug},
  publisher = {American Physical Society},
  doi = {10.1103/PhysRevLett.121.090502},
  url = {https://link.aps.org/doi/10.1103/PhysRevLett.121.090502}
}

@article{Muller2015,
  title = {Interacting two-level defects as sources of fluctuating high-frequency noise in superconducting circuits},
  author = {M\"uller, Clemens and Lisenfeld, J\"urgen and Shnirman, Alexander and Poletto, Stefano},
  journal = {Phys. Rev. B},
  volume = {92},
  issue = {3},
  pages = {035442},
  numpages = {13},
  year = {2015},
  month = {Jul},
  publisher = {American Physical Society},
  doi = {10.1103/PhysRevB.92.035442},
  url = {https://link.aps.org/doi/10.1103/PhysRevB.92.035442}
}

@article{Martinis2022,
  title = {Surface loss calculations and design of a superconducting transmon qubit with tapered wiring},
  author = {Martinis, John M.},
  journal = {npj Quantum Information},
  volume = {8},
  issue = {1},
  pages = {},
  numpages = {},
  year = {2022},
  month = {12},
  publisher = {Nature Research},
  doi = {10.1038/s41534-022-00530-6},
  url = {https://www.nature.com/articles/s41534-022-00530-6}
}

@book{Phillips1981,
  title = {Amorphous Solids},
  author = {W. A. Phillips},
  journal = {},
  volume = {},
  issue = {},
  pages = {},
  numpages = {},
  year = {1981},
  month = {},
  publisher = {
      Springer-Verlag Berlin Heidelberg},
  doi = {},
  url = {}
}

@article{Molina-Ruiz2021,
  title = {Origin of mechanical and dielectric losses from two-level systems in amorphous silicon},
  author = {Molina-Ruiz, M. and Rosen, Y. J. and Jacks, H. C. and Abernathy, M. R. and Metcalf, T. H. and Liu, X. and DuBois, J. L. and Hellman, F.},
  journal = {Phys. Rev. Mater.},
  volume = {5},
  issue = {3},
  pages = {035601},
  numpages = {8},
  year = {2021},
  month = {Mar},
  publisher = {American Physical Society},
  doi = {10.1103/PhysRevMaterials.5.035601},
  url = {https://link.aps.org/doi/10.1103/PhysRevMaterials.5.035601}
}

@article{Osman2023,
  title = {Mitigation of frequency collisions in superconducting quantum processors},
  author = {Osman, Amr and Fern\'andez-Pend\'as, Jorge and Warren, Christopher and Kosen, Sandoko and Scigliuzzo, Marco and Frisk Kockum, Anton and Tancredi, Giovanna and Fadavi Roudsari, Anita and Bylander, Jonas},
  journal = {Phys. Rev. Res.},
  volume = {5},
  issue = {4},
  pages = {043001},
  numpages = {17},
  year = {2023},
  month = {Oct},
  publisher = {American Physical Society},
  doi = {10.1103/PhysRevResearch.5.043001},
  url = {https://link.aps.org/doi/10.1103/PhysRevResearch.5.043001}
}

@article{Lisenfeld2015,
  title = {Observation of directly interacting coherent two-level systems in an amorphous material},
  author = {Lisenfeld, Jürgen and Grabovskij, Grigorij J. and Müller, Clemens and Cole, Jared H. and Weiss, Georg and Ustinov, Alexey V.},
  journal = {Nature Communications},
  volume = {6},
  issue = {1},
  pages = {},
  numpages = {},
  year = {2015},
  month = {Feb},
  publisher = {Nature},
  doi = {10.1038/ncomms7182},
  url = {https://www.nature.com/articles/ncomms7182}
}

@article{Katz2010,
  title = {Lifetime and Coherence of Two-Level Defects in a Josephson Junction},
  author = {Shalibo, Yoni and Rofe, Ya'ara and Shwa, David and Zeides, Felix and Neeley, Matthew and Martinis, John M. and Katz, Nadav},
  journal = {Phys. Rev. Lett.},
  volume = {105},
  issue = {17},
  pages = {177001},
  numpages = {4},
  year = {2010},
  month = {Oct},
  publisher = {American Physical Society},
  doi = {10.1103/PhysRevLett.105.177001},
  url = {https://link.aps.org/doi/10.1103/PhysRevLett.105.177001}
}

@article{Barends2013,
  title = {Coherent Josephson Qubit Suitable for Scalable Quantum Integrated Circuits},
  author = {Barends, R. and Kelly, J. and Megrant, A. and Sank, D. and Jeffrey, E. and Chen, Y. and Yin, Y. and Chiaro, B. and Mutus, J. and Neill, C. and O'Malley, P. and Roushan, P. and Wenner, J. and White, T. C. and Cleland, A. N. and Martinis, John M.},
  journal = {Phys. Rev. Lett.},
  volume = {111},
  issue = {8},
  pages = {080502},
  numpages = {5},
  year = {2013},
  month = {Aug},
  publisher = {American Physical Society},
  doi = {10.1103/PhysRevLett.111.080502},
  url = {https://link.aps.org/doi/10.1103/PhysRevLett.111.080502}
}

@article{GoogleQuantum2024,
  title = {Quantum error correction below the surface code threshold},
  author = {Google Quantum AI and Collaborators},
  journal = {Nature},
  volume = {638},
  issue = {8052},
  pages = {},
  numpages = {},
  year = {2024},
  month = {Dec},
  publisher = {},
  doi = {10.1038/s41586-024-08449-y},
  url = {https://www.nature.com/articles/s41586-024-08449-y}
}

@article{Gao2025,
  title = {Establishing a New Benchmark in Quantum Computational Advantage with 105-qubit Zuchongzhi 3.0 Processor},
  author = {Gao, Dongxin and Fan, Daojin and Zha, Chen and Bei, Jiahao and Cai, Guoqing and Cai, Jianbin and Cao, Sirui and Chen, Fusheng and Chen, Jiang and Chen, Kefu and Chen, Xiawei and Chen, Xiqing and Chen, Zhe and Chen, Zhiyuan and Chen, Zihua and Chu, Wenhao and Deng, Hui and Deng, Zhibin and Ding, Pei and Ding, Xun and Ding, Zhuzhengqi and Dong, Shuai and Dong, Yupeng and Fan, Bo and Fu, Yuanhao and Gao, Song and Ge, Lei and Gong, Ming and Gui, Jiacheng and Guo, Cheng and Guo, Shaojun and Guo, Xiaoyang and Han, Lianchen and He, Tan and Hong, Linyin and Hu, Yisen and Huang, He-Liang and Huo, Yong-Heng and Jiang, Tao and Jiang, Zuokai and Jin, Honghong and Leng, Yunxiang and Li, Dayu and Li, Dongdong and Li, Fangyu and Li, Jiaqi and Li, Jinjin and Li, Junyan and Li, Junyun and Li, Na and Li, Shaowei and Li, Wei and Li, Yuhuai and Li, Yuan and Liang, Futian and Liang, Xuelian and Liao, Nanxing and Lin, Jin and Lin, Weiping and Liu, Dailin and Liu, Hongxiu and Liu, Maliang and Liu, Xinyu and Liu, Xuemeng and Liu, Yancheng and Lou, Haoxin and Ma, Yuwei and Meng, Lingxin and Mou, Hao and Nan, Kailiang and Nie, Binghan and Nie, Meijuan and Ning, Jie and Niu, Le and Peng, Wenyi and Qian, Haoran and Rong, Hao and Rong, Tao and Shen, Huiyan and Shen, Qiong and Su, Hong and Su, Feifan and Sun, Chenyin and Sun, Liangchao and Sun, Tianzuo and Sun, Yingxiu and Tan, Yimeng and Tan, Jun and Tang, Longyue and Tu, Wenbing and Wan, Cai and Wang, Jiafei and Wang, Biao and Wang, Chang and Wang, Chen and Wang, Chu and Wang, Jian and Wang, Liangyuan and Wang, Rui and Wang, Shengtao and Wang, Xiaomin and Wang, Xinzhe and Wang, Xunxun and Wang, Yeru and Wei, Zuolin and Wei, Jiazhou and Wu, Dachao and Wu, Gang and Wu, Jin and Wu, Shengjie and Wu, Yulin and Xie, Shiyong and Xin, Lianjie and Xu, Yu and Xue, Chun and Yan, Kai and Yang, Weifeng and Yang, Xinpeng and Yang, Yang and Ye, Yangsen and Ye, Zhenping and Ying, Chong and Yu, Jiale and Yu, Qinjing and Yu, Wenhu and Zeng, Xiangdong and Zhan, Shaoyu and Zhang, Feifei and Zhang, Haibin and Zhang, Kaili and Zhang, Pan and Zhang, Wen and Zhang, Yiming and Zhang, Yongzhuo and Zhang, Lixiang and Zhao, Guming and Zhao, Peng and Zhao, Xianhe and Zhao, Xintao and Zhao, Youwei and Zhao, Zhong and Zheng, Luyuan and Zhou, Fei and Zhou, Liang and Zhou, Na and Zhou, Naibin and Zhou, Shifeng and Zhou, Shuang and Zhou, Zhengxiao and Zhu, Chengjun and Zhu, Qingling and Zou, Guihong and Zou, Haonan and Zhang, Qiang and Lu, Chao-Yang and Peng, Cheng-Zhi and Zhu, Xiaobo and Pan, Jian-Wei},
  journal = {Phys. Rev. Lett.},
  volume = {134},
  issue = {9},
  pages = {090601},
  numpages = {7},
  year = {2025},
  month = {Mar},
  publisher = {American Physical Society},
  doi = {10.1103/PhysRevLett.134.090601},
  url = {https://link.aps.org/doi/10.1103/PhysRevLett.134.090601}
}

@article{Riken2024,
  title = {Realization of High-Fidelity CZ Gate Based on a Double-Transmon Coupler},
  author = {Li, Rui and Kubo, Kentaro and Ho, Yinghao and Yan, Zhiguang and Nakamura, Yasunobu and Goto, Hayato},
  journal = {Phys. Rev. X},
  volume = {14},
  issue = {4},
  pages = {041050},
  numpages = {30},
  year = {2024},
  month = {Nov},
  publisher = {American Physical Society},
  doi = {10.1103/PhysRevX.14.041050},
  url = {https://link.aps.org/doi/10.1103/PhysRevX.14.041050}
}

@article{Fowler2012,
  title = {Towards Practical Classical Processing for the Surface Code},
  author = {Fowler, Austin G. and Whiteside, Adam C. and Hollenberg, Lloyd C. L.},
  journal = {Phys. Rev. Lett.},
  volume = {108},
  issue = {18},
  pages = {180501},
  numpages = {4},
  year = {2012},
  month = {May},
  publisher = {American Physical Society},
  doi = {10.1103/PhysRevLett.108.180501},
  url = {https://link.aps.org/doi/10.1103/PhysRevLett.108.180501}
}

@article{FowlerMartinis2012,
  title = {Surface codes: Towards practical large-scale quantum computation},
  author = {Fowler, Austin G. and Mariantoni, Matteo and Martinis, John M. and Cleland, Andrew N.},
  journal = {Phys. Rev. A},
  volume = {86},
  issue = {3},
  pages = {032324},
  numpages = {48},
  year = {2012},
  month = {Sep},
  publisher = {American Physical Society},
  doi = {10.1103/PhysRevA.86.032324},
  url = {https://link.aps.org/doi/10.1103/PhysRevA.86.032324}
}

@article{positionPaper2024,
  title = {How to Build a Quantum Supercomputer: Scaling from Hundreds to Millions of Qubits},
  author = {Mohseni, Masoud. and Scherer, Artur. and Johnson, K. Grace and Wertheim, Oded. and Otten, Matthew and Aadit, Navid Anjum and Alexeev, Yuri and Bresniker, Kirk M. and Camsari, Kerem Y. and Chapman, Barbara and Martinis, John M.},
  year = {2024},
  month = {Nov},
  journal = {arXiv preprint arXiv:2411.10406},
  archivePrefix = {arXiv},
  primaryClass = {quant-ph},
  doi = {10.48550/arXiv.2411.10406},
  url = {https://arxiv.org/abs/2411.10406}
}

@article{Bylander2019,
  title = {Decoherence benchmarking of superconducting qubits},
  author = {Burnett, Jonathan J. and Bengtsson, Andreas and Scigliuzzo, Marco and Niepce, David and Kudra, Marina and Delsing, Per and Bylander, Jonas},
  journal = {npj Quantum Information},
  volume = {5},
  issue = {1},
  pages = {},
  numpages = {},
  year = {2019},
  month = {Jun},
  publisher = {Nature Portfolio},
  doi = {10.1038/s41534-019-0168-5},
  url = {https://www.nature.com/articles/s41534-019-0168-5}
}

@article{Weides2019,
  title = {Correlating Decoherence in Transmon Qubits: Low Frequency Noise by Single Fluctuators},
  author = {Schl\"or, Steffen and Lisenfeld, J\"urgen and M\"uller, Clemens and Bilmes, Alexander and Schneider, Andre and Pappas, David P. and Ustinov, Alexey V. and Weides, Martin},
  journal = {Phys. Rev. Lett.},
  volume = {123},
  issue = {19},
  pages = {190502},
  numpages = {6},
  year = {2019},
  month = {Nov},
  publisher = {American Physical Society},
  doi = {10.1103/PhysRevLett.123.190502},
  url = {https://link.aps.org/doi/10.1103/PhysRevLett.123.190502}
}

@article{Macha2010,
  title = {Losses in coplanar waveguide resonators at millikelvin temperatures},
  author = {Macha, P. and van der Ploeg, S. H. W. and Oelsner, G. and Il'ichev, E. and Meyer, H-G. and Wünsch, S. and Siegel, M.},
  journal = {Appl. Phys. Lett.},
  volume = {96},
  issue = {6},
  pages = {},
  numpages = {},
  year = {2010},
  month = {Feb},
  publisher = {American Institute of Physics},
  doi = {10.1063/1.3309754},
  url = {https://pubs.aip.org/aip/apl/article/96/6/062503/167025/Losses-in-coplanar-waveguide-resonators-at}
}

@article{Lisenfeld2019,
  title = {Electric field spectroscopy of material defects in transmon qubits},
  author = {Lisenfeld, Jürgen and Bilmes, Alexander and Megrant, Anthony and Barends, Rami and Kelly, Julian and Klimov, Paul and Weiss, Georg and Martinis, John M. and Ustinov, Alexey V.},
  journal = {npj Quantum Information},
  volume = {5},
  issue = {1},
  pages = {},
  numpages = {},
  year = {2019},
  month = {Nov},
  publisher = {Nature Portfolio},
  doi = {10.1038/s41534-019-0224-1},
  url = {https://www.nature.com/articles/s41534-019-0224-1}
}

@article{Schickfus1977,
  title = {Saturation of the dielectric absorption of vitreous silica at low temperatures},
  author = {Schickfus, M. Von and Hunklinger, S.},
  journal = {Phys. Lett. A},
  volume = {64},
  issue = {1},
  pages = {144-146},
  numpages = {},
  year = {1977},
  month = {Nov},
  publisher = {Elsevier},
  doi = {10.1016/0375-9601(77)90558-8},
  url = {https://www.sciencedirect.com/science/article/pii/0375960177905588}
}

@article{Faoro20151,
  title = {Interacting tunneling model for two-level systems in amorphous materials and its predictions for their dephasing and noise in superconducting microresonators},
  author = {Faoro, Lara and Ioffe, Lev B.},
  journal = {Phys. Rev. B},
  volume = {91},
  issue = {1},
  pages = {014201},
  numpages = {11},
  year = {2015},
  month = {Jan},
  publisher = {American Physical Society},
  doi = {10.1103/PhysRevB.91.014201},
  url = {https://link.aps.org/doi/10.1103/PhysRevB.91.014201}
}

@ARTICLE{Murray2018,
  author={Murray, Conal E. and Gambetta, Jay M. and McClure, Douglas T. and Steffen, Matthias},
  journal={IEEE Transactions on Microwave Theory and Techniques}, 
  title={Analytical Determination of Participation in Superconducting Coplanar Architectures}, 
  year={2018},
  volume={66},
  number={8},
  pages={3724-3733},
  doi={10.1109/TMTT.2018.2841829},
  url = {https://ieeexplore.ieee.org/document/8386844}
}

@ARTICLE{Bilmes2021,
  author={Alexander Bilmes and Alexander K Händel and Serhii Volosheniuk and Alexey V Ustinov and Jürgen Lisenfeld},
  journal={Supercond. Sci. Technol.}, 
  title={In-situ bandaged Josephson junctions for superconducting quantum processors}, 
  year={2021},
  volume={34},
  number={12},
  pages={},
  doi={10.1088/1361-6668/ac2a6d},
  url = {https://iopscience.iop.org/article/10.1088/1361-6668/ac2a6d}
}

@article{Thorbeck2023,
  title = {Two-Level-System Dynamics in a Superconducting Qubit Due to Background Ionizing Radiation},
  author = {Thorbeck, Ted and Eddins, Andrew and Lauer, Isaac and McClure, Douglas T. and Carroll, Malcolm},
  journal = {PRX Quantum},
  volume = {4},
  issue = {2},
  pages = {020356},
  numpages = {14},
  year = {2023},
  month = {Jun},
  publisher = {American Physical Society},
  doi = {10.1103/PRXQuantum.4.020356},
  url = {https://link.aps.org/doi/10.1103/PRXQuantum.4.020356}
}

@misc{tlsDephasing,
  note = {The defect relaxation rate includes contributions from dephasing and loss: $\Gamma_{\rm d} = \Gamma_{\phi,\rm d} + \Gamma_{1,\rm d}/2$. We neglect the baseline qubit relaxation rate as it is small compared to the defect relaxation rate.}
}

@article{BrojenyClem2003,
  title = {Magnetic-field and current-density distributions in thin-film superconducting rings and disks},
  author = {Babaei Brojeny, Ali A. and Clem, John R.},
  journal = {Phys. Rev. B},
  volume = {68},
  issue = {17},
  pages = {174514},
  numpages = {9},
  year = {2003},
  month = {Nov},
  publisher = {American Physical Society},
  doi = {10.1103/PhysRevB.68.174514},
  url = {https://link.aps.org/doi/10.1103/PhysRevB.68.174514}
}

@article{Rafferty2021,
  title = {Spurious Antenna Modes of the Transmon Qubit},
  author = {O. Rafferty and S. Patel and C. H. Liu and S. Abdullah and C. D. Wilen and D. C. Harrison and R. McDermott},
  journal = {arXiv preprint arXiv:2103.06803},
  archivePrefix = {arXiv},
  primaryClass = {quant-ph},
  year = {2021},
  month = {Mar},
  publisher = {},
  doi = {10.48550/arXiv.2103.06803},
  url = {https://arxiv.org/abs/2103.06803}
}

@article{You2022,
  title = {Stabilizing and Improving Qubit Coherence by Engineering the Noise Spectrum of Two-Level Systems},
  author = {You, Xinyuan and Huang, Ziwen and Alyanak, Ugur and Romanenko, Alexander and Grassellino, Anna and Zhu, Shaojiang},
  journal = {Phys. Rev. Appl.},
  volume = {18},
  issue = {4},
  pages = {044026},
  numpages = {15},
  year = {2022},
  month = {Oct},
  publisher = {American Physical Society},
  doi = {10.1103/PhysRevApplied.18.044026},
  url = {https://link.aps.org/doi/10.1103/PhysRevApplied.18.044026}
}

@article{Bal2024,
    title={Systematic improvements in transmon qubit coherence enabled by niobium surface encapsulation},
    author = {Bal, Mustafa and Murthy, Akshav A. and Zhu, Shaojiang and Crisa, Francesco and You, Xinyuan and Huang, Ziwen and Roy, Tanay and Lee, Jaeyel and van Zanten, David and Pilipenko, Roman and Nekrashevich, Ivan and Lunin, Andrei and Bafia, Daniel, and Grassellino, Anna},
    journal = {npj Quantum Information},
    volume = {10},
    issue = {1},
    year = {2024},
    mon = {Apr},
    doi = {10.1038/s41534-024-00840-x},
    url = {https://www.nature.com/articles/s41534-024-00840-x}
}

@unpublished{YuCarruzco2025,
    author = {Yu, Clare C. and Carruzzo, Herve.},
    note = {Unpublished},
}

@article{Lisenfeld2025,
  title = {Mapping the positions of Two-Level-Systems on the surface of a superconducting transmon qubit},
  author = {Lisenfeld, Jürgen and Händel, Alexander K and Daum, Etienne and Berlitz, Benedikt and Bilmes, Alexander and Ustinov, Alexey V.},
  journal = {arXiv preprint arXiv:2511.05365},
  archivePrefix = {arXiv},
  primaryClass = {quant-ph},
  year = {2025},
  month = {Nov},
  publisher = {},
  doi = {10.48550/arXiv.2511.05365},
  url = {https://arxiv.org/abs/2511.05365}
}

@article{rawData,
    title = {Data for ``Statistics of Strongly Coupled Defects in Superconducting Qubits"},
    author={S. Weeden},
    journal={Mendeley Data},
    year = {2026},
    version={V1},
    doi = {10.17632/zk2mcxvkhy.1},
    url={https://data.mendeley.com/datasets/zk2mcxvkhy/1}
}

\end{document}